\documentclass[aps,prb,noshowpacs,twocolumn,showpacs,superscriptaddress,amsmath,amssymb]{revtex4-1}
\usepackage{amsmath}
\usepackage{amssymb}
\usepackage{amsthm}

\usepackage{amsbsy}
\usepackage{amstext}
\usepackage{graphicx}
\usepackage{color}
\usepackage{float}
\usepackage{siunitx}
\usepackage{wasysym}

\usepackage{mathtools}
\usepackage{subfigure}
\makeatletter
\usepackage{amsfonts}
\usepackage{dcolumn}
\usepackage{bbold,bm}

\usepackage{makecell}

\def\dd{\mathrm{d}}
\def\ss{\mathbf{s}}

\def\dd{\boldsymbol{\delta}}

\makeatother

\begin{document}

\title{Phonon-mediated dimensional crossover in bilayer CrI$_3$}
\author{M. Rodriguez-Vega}
\thanks{These two authors contributed equally}
\affiliation{Department of Physics, The University of Texas at Austin, Austin, TX 78712, USA}
\affiliation{Department of Physics, Northeastern University, Boston, MA 02115, USA}

\author{Ze-Xun Lin}
\thanks{These two authors contributed equally}
\affiliation{Department of Physics, The University of Texas at Austin, Austin, TX 78712, USA}
\affiliation{Department of Physics, Northeastern University, Boston, MA 02115, USA}

\author{A. Leonardo}
\affiliation{Donostia International Physics Center, Paseo Manuel de Lardizabal 4, 20018 San Sebastian, Spain}
\affiliation{Applied Physics Department II, University of the Basque Country UPV/
EHU, Bilbao, Spain.}

\author{A. Ernst}
\affiliation{Institut f\"ur Theoretische Physik, Johannes Kepler Universit\"at, A 4040 Linz, Austria}
\affiliation{Max-Planck-Institut f\"ur Mikrostrukturphysik, Weinberg 2, D-06120 Halle, Germany}

\author{G. Chaudhary}
\affiliation{James Franck Institute, Department of Physics, The University of Chicago, Chicago, IL 60637}
\author{M. G. Vergniory}
\affiliation{Donostia International Physics Center, Paseo Manuel de Lardizabal 4, 20018 San Sebastian, Spain}
\affiliation{IKERBASQUE, Basque Foundation for Science, Bilbao, Spain}

\author{Gregory A. Fiete}
\affiliation{Department of Physics, Northeastern University, Boston, MA 02115, USA}
\affiliation{Department of Physics, The University of Texas at Austin, Austin, TX 78712, USA}
\affiliation{Department of Physics, Massachusetts Institute of Technology, Cambridge, MA 02139, USA}

\begin{abstract}
In bilayer CrI$_3$, experimental and theoretical studies suggest that the magnetic order is closely related to the layer staking configuration. In this work, we study the effect of dynamical lattice distortions, induced by non-linear phonon coupling, in the magnetic order of the bilayer system. We use density functional theory to determine the phonon properties and group theory to obtain the allowed phonon-phonon interactions. We find that the bilayer structure possesses low-frequency Raman modes that can be non-linearly activated upon the coherent photo-excitation of a suitable infrared phonon mode. This transient lattice modification in turn inverts the sign of the interlayer spin interaction for parameters accessible in experiments, indicating a low-frequency light-induced antiferromagnet-to-ferromagnet transition.
\end{abstract}

\date{\today}
\maketitle
%\tableofcontents

The control of ordered states of matter such as magnetism, superconductivity or charge and spin density waves is one of the more sought after effects in the field. In equilibrium, this can be achieved by turning the knobs provided by temperature, strain, pressure, or chemical composition. However, the nature of these methods limits the possibility to integrate the materials into devices for technological applications due to undesirably slow control and non-reversibility. In recent years, a new approach has emerged which allows \textit{in-situ} manipulation: driving systems out of equilibrium by irradiating them with light~\cite{oka2009, lindner2011,FORST201324,subedi2014, Mentink2015, gu2017, juraschek2017,Juraschek2017b,subedi2017,Babadi2017,liu2018,Juraschek2018,Juraschek2019,Juraschek2020,juraschek2019phonomagnetic,kalsha2018,gu2018,fechner2016,Hejazi2019,Sentef2016,Sentef2017,Tancogne2018,chaudhary2019phononinduced,Chaudhary2019,vogl2020effective, vogl2020floquet, asmar2020floquet, ke2020nonequilibrium, PhysRevX.9.021037,baldini2020,PhysRevLett.107.216601,PhysRevLett.116.016802,PhysRevMaterials.2.064401}. Recent experiments have demonstrated the existence of Floquet states in topological insulators~\cite{Wang453,Mahmood2016}, the possibility to transiently enhance superconductivity~\cite{Fausti189,Mankowsky2014,mitrano2016}, the existence of light-induced anomalous Hall states in graphene~\cite{mciver2018lightinduced}, light-induced metastable charge-density-wave states in 1T-TaS$_2$~\cite{Vaskivskyie1500168}, optical pulse-induced metastable metallic phases hidden in charge ordered insulating phases ~\cite{Zhang2016, teitelbaum2019dynamics}, and metastable ferroelectric phases in titanates~\cite{Nova1075}.

\begin{figure}[t]
	\begin{center}
		\subfigure{\includegraphics[width=8.50cm]{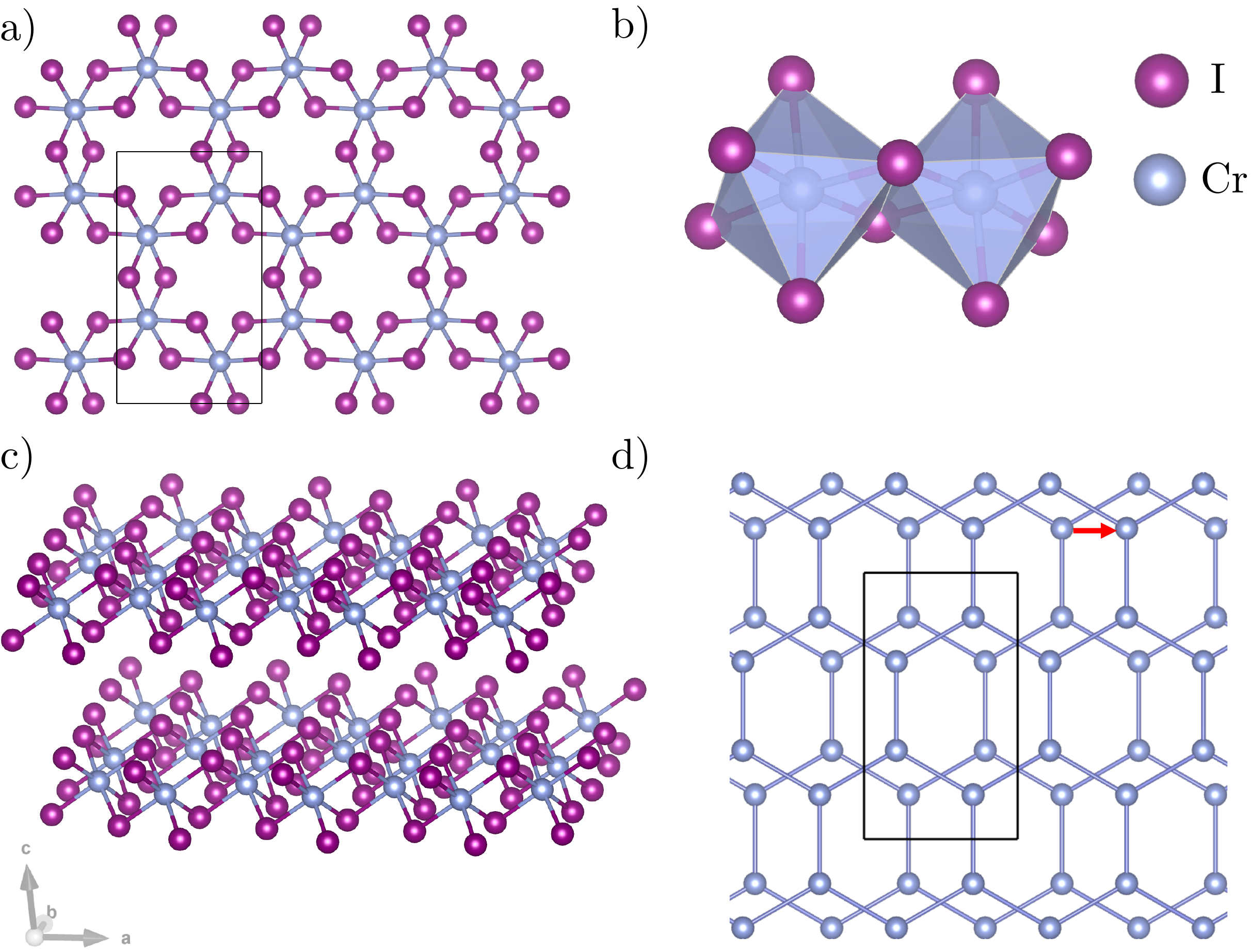}}		
		\caption{(Color online) a) Monolayer CrI$_3$ lattice structure. The conventional unit cell is shown with solid lines. b) Cr$^{+3}$ atoms surrounded by an edge-sharing I octahedral cage. c) b-CrI$_3$ crystal structure with space group $C2/m$ associated with the AFM ground state. The top layer is shifted with respect to the bottom layer by $[1/3\;0\;0]$. d) Top view with I atoms suppressed for clarity. The red arrow indicates the relative shift. The black box indicates the conventional unit cell. The lattice structures where created with VESTA~\cite{Momma:db5098}.}
		\label{fig:cri3}
	\end{center}
\end{figure}

Finding suitable platforms to realize non-equilibrium transitions represents the first main challenge. Recently, interest in the van der Waals bulk ferromagnet chromium triiodide (CrI$_3$)~\cite{Dillon1965,mcguire2015} has been renewed with the discovery that it is stable in its monolayer form, where the chromium atoms arrange in a hexagonal lattice and the iodine atoms order on a side-sharing octahedral cage around each chromium atom as shown in Fig. \ref{fig:cri3}(a-b). Monolayer CrI$_3$ presents out-of-plane magnetization stabilized by anisotropies~\cite{mermin1966} and a Curie temperature $T \sim 45$ K~\cite{huang2017}. The origin of the anisotropies is still a subject of intense theoretical and experimental investigations~\cite{chen2018,lee2020fundamental,lado2017}. 

In bulk form, CrI$_3$ exhibits a structural phase transition near $T= 210-220 $~K. This structural transition is accompanied by an anomaly in the magnetic susceptibility, but no magnetic ordering~\cite{mcguire2015}. At $T=61$ K, CrI$_3$ exhibits a transition from paramagnet to ferromagnet~\cite{mcguire2015}, with an easy-axis perpendicular to the 2D planes. Evidence suggests that CrI$_3$ is a Mott insulator with a band gap close to 1.2 eV~\cite{Dillon1965,mcguire2015}. Recent experiments have measured large tunneling magnetoresistance ~\cite{wangzhe2018,song2018}, suggesting potential applications in spintronics devices. 	

Bilayer CrI$_3$ (b-CrI$_3$) presents an antiferromagnetic (AFM) groundstate~\cite{huang2017,seyler2018,klein1218,song2018,sun2019}, with monoclinic crystal structure (Fig. \ref{fig:cri3}(c-d)). Single-spin microscopy~\cite{Thiel973} and polarization resolved Raman spectroscopy~\cite{ubrig2020} measurements have established a strong connection between the magnetic order and the stacking configuration in few-layers CrI$_3$. Furthermore, it has been shown that the magnetic order can be controlled in equilibrium by doping ~\cite{huang2018} and applying pressure ~\cite{song2019pressure} to b-CrI$_3$. These results have been accompanied by theoretical studies, which find that the AFM order is linked to the lattice configuration~\cite{sivadas2018,jang2019,soriano2019,soriano2020}. In particular, orbital-dependent magnetic force calculations show that the stacking pattern can suppress or enhance the $e_g-t_{2g}$ interaction and correspondingly favor a AFM or FM order~\cite{jang2019}.

In this Letter, we leverage these theoretical and experimental results in equilibrium, and consider the possibility to dynamically tune the magnetic order in b-CrI$_3$ using low-frequency light to coherently drive suitable phonon modes. We start with a group theory analysis to determine the feasibility of the non-linear phonon process required. Then, we perform first principles calculations to find phonon frequencies, eigenmodes and non-linear phonon coupling strengths. We then analyze the equations of motion for the driven phonons and their impact on the lattice structure. Finally, we determine the effect of such transient lattice deformations on the magnetic order and find the possibility to induce a sign change in the interlayer exchange interaction using experimentally-accessible parameters. 

\textit{Group theory analysis.} Recent first-principles studies indicate that there is a direct relation between the magnetic ground state and the relative stacking order between the layers~\cite{sivadas2018,jiang2019,soriano2019}. The FM phase presents an AB stacking with space group R$\bar 3$ (point group S$_6$), while the AFM ground state is accompanied by an AB' stacking with space group C2/m  and point group C$_{2h}$~\cite{sivadas2018} (Fig. \ref{fig:cri3}(d)). AFM and FM structures are related by a relative shift of the layers leaving each individual layer unaltered. Since experiments find AFM order in the ground state~\cite{huang2017}, in our analysis we assume the configuration corresponding to the C$_{2h}$ space group. The primitive unit cell contains $4$ Cr and $12$ I atoms, for a total of  $N=16$ atoms. The conventional to primitive unit cell transformation, and the C$_{2h}$ point group character table are listed in~\cite{supplemental}. The total number of phonon modes is then $3 N  = 48$. We obtain that the equivalence representation is given by 
$%\begin{equation}
\Gamma^{equiv} =  5 A_g \oplus 3 B_g \oplus 3 A_u \oplus 5 B_u
$. %\end{equation} 
In the $C_{2h}$ point group, the representation of the vector is $\Gamma_{vec} = 2A_u \oplus B_u$, which leads to the lattice vibration representation 
$%\begin{equation}
\Gamma_{latt. vib.} = \Gamma^{equiv} \otimes \Gamma_{vec}  =  13 A_g \oplus 11 B_g \oplus 11 A_u \oplus 13 B_u
$. %\end{equation}
From the symmetry of the generating functions (see the Supplemental Material \cite{supplemental}), 24 modes are Raman active (13 with totally symmetric $A_g$ representation and 11 with $B_g$ representation) and 24 infrared active modes~\cite{kroumova2003}. 

\begin{figure}[t]
	\begin{center}
		\includegraphics[width=8.5cm]{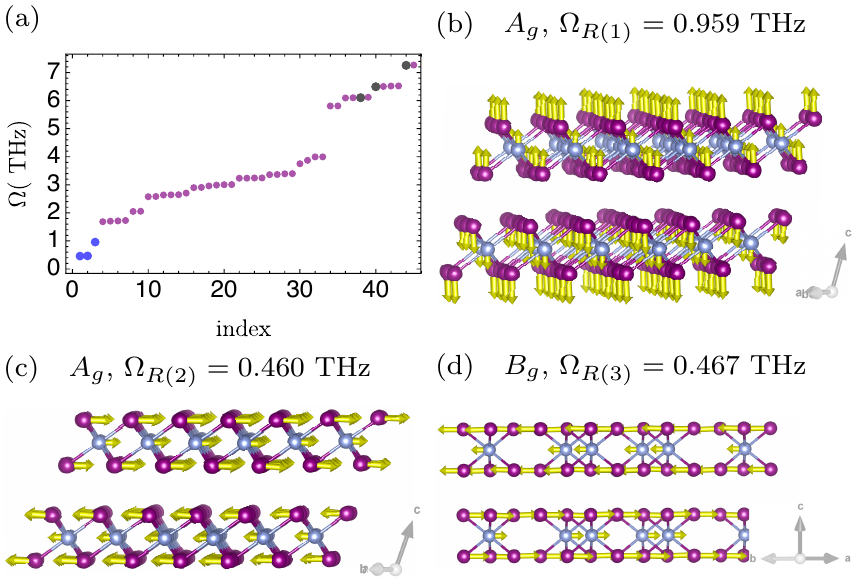}
		\caption{(Color online) (a) b-CrI$_3$ phonon frequencies in the space group $C2/m$ and AFM  ground state. The blue dots correspond to Raman modes involving layer relative shifts. The gray dots correspond to IR modes that directly couple to the light pulse. The acoustic modes are not shown. (b-d) b-CrI$_3$ low-frequency Raman phonon displacements relevant for the non-linear dynamics.}
		\label{fig:bilayer_dft_results}
	\end{center}
\end{figure}

Here, we posit that a Raman mode involving a relative shift between the layers might influence the magnetic order. In order to test if such a mode is allowed by symmetry, we construct the projection operators~\cite{GroupTheoryDress2008, gtpack1, gtpack2} 
$
\hat P^{(\Gamma_n)}_{kl} = \frac{l_n}{h} \sum_{C_\alpha} \left( D_{kl}^{(\Gamma_n)}(C_\alpha) \right)^* \hat P(C_\alpha)$,
where $\Gamma_n$ are the irreducible representations, $C_\alpha$ are the elements of the group, $D_{kl}^{(\Gamma_n)}(C_\alpha) $ is the irreducible matrix representation of element $C_\alpha$, $h$ is the order of the group, and $l_n$ is the dimension of the irreducible representation. Finally, $\hat P(C_\alpha)$ are $3N\times 3N$ matrices that form the displacement representation. Applying the projection operators $\hat P^{A_g}$ and $\hat P^{B_g}$ to random displacements of the atoms, we find that modes with one layer uniformly displaced in the $[1 \; 1 \; 0]$  direction, while the other in the $[\bar1 \; \bar1 \; 0]$ direction is allowed by symmetry and belong to the totally-symmetric $A_g$ representation (Fig. \ref{fig:bilayer_dft_results}(c)). Similarly, modes where one layer is displaced in the $[0 \; 0 \; 1]$ direction and the other one in the $[0 \; 0 \; \bar 1]$ belongs to the $A_g$ representation (Fig. \ref{fig:bilayer_dft_results}(b)). On the other hand, layer displacements in the directions $[\bar 1 \; 1 \; 0]$ and $[\;1 \; \bar 1 0]$, belong to the $B_g$ representation (Fig. \ref{fig:bilayer_dft_results}(d)). We will show that these Raman modes can be manipulated via indirect coupling with light to control the magnetic order.

\textit{Phonons.} Once we determine that relative-shift modes are allowed by symmetry, we calculate the phonon frequencies using density functional perturbation theory (DFPT) and finite difference methods as implemented in QUANTUM ESPRESSO \cite{Giannozzi_2009,Giannozzi_2017} and VASP \cite{kresse1996,kresse1996b}, respectively. We find excellent agreement among all the approaches considered (see \cite{supplemental} for details).  

In Fig.~\ref{fig:bilayer_dft_results}(a), we plot the full set of frequencies of the $\Gamma$-point phonons.  We find that the three low-frequency modes (apart from the three omitted zero-frequency acoustic modes) are Raman active, and correspond to relative displacement between the layers in different directions, in agreement with the group theory results. The lowest-frequency mode, $\Omega=0.460$~THz, belongs to the $A_g$ representation, and the real-space displacement is shown in Fig.~\ref{fig:bilayer_dft_results}(c). The next phonon mode is very close in frequency, $\Omega=0.467$~THz, however, it belongs to the $B_g$ representation (Fig.~\ref{fig:bilayer_dft_results}(d)). The mode with frequency $\Omega=0.959$~THz belongs to the $A_g$ representation and corresponds to a relative displacement perpendicular to the layers, as shown in Fig.~\ref{fig:bilayer_dft_results}(b). 

%%%%%%%%%%%%%%%%%%%%%%%%%%%%%
% Non-linear phononics
%%%%%%%%%%%%%%%%%%%%%%%%%%%%%
Non-linear phonon processes have been proposed for transient modification of the symmetries of the system, which can be accompanied by changes in the ground-state properties~\cite{FORST201324,subedi2014,juraschek2017,Juraschek2017b,subedi2017,Juraschek2018,Juraschek2019,Juraschek2020,juraschek2019phonomagnetic}. Now, we derive the non-linear phonon potential resulting from coupling  between infrared $(Q_{{\rm IR}})$ and Raman $(Q_{{\rm R}})$ active modes  in b-CrI$_3$. In an invariant polynomial under the operations of a given group, coupling between two modes is allowed only if it contains the totally symmetric representation~\cite{GroupTheoryDress2008,gtpack1,gtpack2}.  In principle, an IR mode is allowed to couple non-linearly to all $A_g$ and $B_g$ Raman modes in the $C_{2h}$ point group. However, as we will show, we can focus on  the modes involving relative motion between the layers because they possess very low frequency, compared with the rest of the Raman phonons.  Up to cubic order, the non-linear potential functional including the three low-frequency phonon modes is given by  
%We consider the coherent light-induced excitation of a non-degenerate infrared active mode with representation $B_{u}$.
%
%
\begin{align}\nonumber
&V[Q_{\text{IR}},Q_{\text{R}(i)}] =  \frac{1}{2}\Omega^2_{\text{IR}} Q_{\text{IR}}^2+ \sum_{i=1}^3 \frac{1}{2}\Omega^2_{\text{R}(i)} Q^{2}_{\text{R}(i)} \\ \nonumber 
&+\sum_{i=1}^2 \frac{ \beta_i}{3} Q_{\text{R(i)}}^3 + Q_{\text{IR}}^2 \sum^2_{i=1} \gamma_i Q_{\text{R(i)}} + \delta Q_{\text{R(1)}}^2 Q_{\text{R(2)}}\\ & + \epsilon Q_{\text{R(1)}} Q_{\text{R(2)}}^2+   Q_{\text{R(3)}}^2  \sum^2_{i=1} \zeta_i Q_{\text{R(i)}} .
\label{eq:non-linear-pot_main}
\end{align}

The numerical value of the coefficients is obtained using first-principles calculations. In the Supplemental Material \cite{supplemental} we outline the procedure we used following Ref.~[\onlinecite{subedi2014}], we plot the energy surfaces obtained by varying the corresponding phonon mode amplitudes, and display the numerical values of the coefficients obtained by fitting Eq. \eqref{eq:non-linear-pot_main}. 

Under an external drive with frequency $\Omega$, the potential acquires the time-dependent term~\cite{baroni2001,forst2011}
$%\begin{align}
V_D[Q_{\text{IR}}]  = \boldsymbol{Z^*} \cdot \boldsymbol{E_0} \sin(\Omega t) F(t) Q_{\text{IR}}
$, %\end{align}
where $\boldsymbol{E_0}$ is the electric field amplitude, and $\boldsymbol{Z^*}$ is the mode effective charge vector~\cite{gonze1997, baroni2001}. $F(t)=\exp\{ -t^2/(2 \tau^2)\}$ is the Gaussian laser profile, with variance $\tau^2$.
Assuming that damping can be neglected, the general differential equations governing the dynamics of one infrared mode coupled to $m$ Raman modes are obtained from the relations $\partial^2_t Q_{\text{R}(i)}   = -\partial_{Q_{\text{R}(i)}} V[Q_{\text{IR}},Q_{\text{R}(i)}]$, for $i=1,\cdots,m$, and
$\partial^2_t Q_{\text{IR}}     = -\partial_{Q_{\text{IR}}} V[Q_{\text{IR}},Q_{\text{R}(i)}]$,
which corresponds to a set of $m+1$ coupled differential equations that we solve numerically in the general case. In the absence of coupling with the Raman modes, the IR mode dynamics are described by 
$
\partial^2_t Q_{\text{IR}}  = -\Omega^2_{\text{IR}} Q_{\text{IR}}-
Z^* E_0 \sin(\Omega t) F(t). 
$
In the resonant case $\Omega = \Omega_{\text{IR}} $, and impulsive limit $\Omega_{\text{IR}} \tau \ll 1$, we find
$%\begin{equation}
Q_{\text{IR}}(t) = \sqrt{2 \pi} Z^* E_0 \tau/\Omega_{\text{IR}} \cos \left(\Omega_{\text{IR}} t \right) \sinh\left[ (\Omega_{\text{IR}} \tau)^2 \right]e^{-(\Omega_{\text{IR}} \tau)^2}
%\label{eq:qir_approx}
$%\end{equation}
with boundary conditions $Q_{\text{IR}}(-\infty)=\partial_t Q_{\text{IR}}(-\infty)=0$~\cite{subedi2014}. The amplitude of the excited IR modes scales linearly with the electric field and the mode effective charge. 

Now we add coupling with one $A_{g}$ Raman mode. The potential in Eq. \eqref{eq:non-linear-pot_main} simplifies to
%
%\begin{align}
$
V[Q_{\text{IR}}, Q_{\text{R}}]  = \frac{1}{2}\Omega^2_{\text{IR}} Q_{\text{IR}}^2+\frac{1}{2}\Omega^2_{\text{R}} Q_{\text{R}}^2 +\gamma Q_{\text{IR}}^2 Q_R.
$
%\end{align}
%
%
The cubic term $\gamma$ is responsible for the \textit{ionic Raman scattering} (IRS) ~\cite{wallis1971,forst2011}. Within this mechanism, the infrared active mode is used to drive Raman scattering processes through anharmonic terms in the potential, and leads to coherent oscillations around a new displaced equilibrium position. Theoretical works have also proposed this cubic non-linear coupling mechanism to tune magnetic order in RTiO$_3$~\cite{kalsha2018,gu2018}, investigate light-induced dynamical symmetry breaking~\cite{subedi2014}, modulate the structure of YBa$_2$Cu$_3$O and related effects in the magnetic order~\cite{fechner2016}. On the experimental side, the response of YBa$_2$Cu$_3$O$_{6+x}$ to optical pulses has been investigated~\cite{mankowsky2015}, and experimental detection of possible light-induced superconductivity has been reported~\cite{mitrano2016}. 

From the equilibrium condition $\partial_{Q_{\text{R}}} V[Q_{\text{IR}}, Q_{\text{R}}] =0$, we find that the potential is minimized when 
$
Q_{\text{R}} = -\gamma Q^2_{\text{IR}}/ \Omega^2_{\text{R}}
$
~\cite{juraschek2017}. Therefore, we obtain larger displaced equilibrium positions effects for low-frequency Raman modes. This argument allows us to limit our discussion to the three low-frequency Raman modes shown in Fig.~\ref{fig:bilayer_dft_results}(b-d). Now, considering the cubic term as a perturbation, we find 
\begin{align} \nonumber
Q_{\text{R}}(t) &= \frac{\gamma \pi (Z^* E_0 \tau^3)^2}{\left(4 \Omega^2_{\text{IR}} - \Omega^2_{\text{R}}\right)} \frac{\Omega^2_{\text{IR}}}{\Omega^2_{\text{R}}} \left(\right. \Omega^2_{\text{R}} \cos(2\Omega_{\text{IR}}t ) +\\
& 2 (\Omega^2_{\text{IR}}-\Omega^2_{\text{R}}) \cos(\Omega_{\text{R}}t ) + \Omega^2_{\text{R}}-4\Omega^2_{\text{IR}} \left. \right).
\end{align}
In the resonant limit $\Omega_{\text{R}} = 2 \Omega_{\text{IR}}$, the solution is given by
$
Q_{\text{R}}(t) =-\gamma \pi (Z^* E_0 \tau^3)^2 \sin(\Omega_{\text{IR}}t ) \left(  \sin(\Omega_{\text{IR}}t ) + \Omega_{\text{IR}}t \cos(\Omega_{\text{IR}}t ) \right)/2.
$

Additional constrains for the IR mode selection arise from current experimental capabilities for strong THz pulse generation. Strong fields of up to $100$~MVcm$^{-1}$ have been achieved in the literature in the range $15-50$ THz~\cite{sell2008,kampfrath2013}.

\begin{figure}[t]
	\begin{center}
		\includegraphics[width=8.5cm]{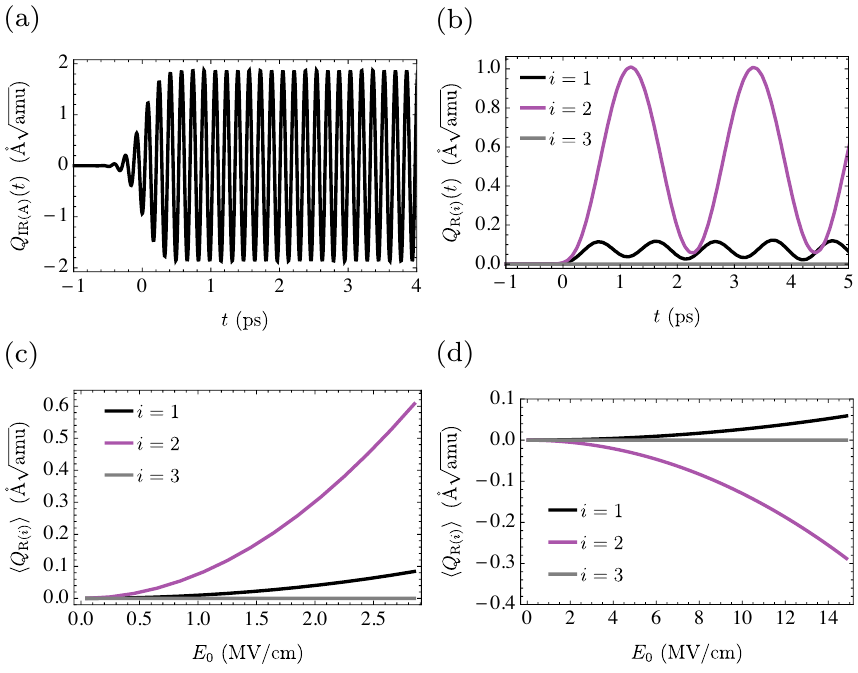}
		\caption{(Color online) 
		(a) Infrared $Q_{\text{IR}(A)}$ and (b) Raman modes $Q_{\text{R}(i)}$ oscillations as a function of time for b-CrI$_3$ for a laser incident in the $y$-direction with $E = 4$ MV/cm, $\tau=0.2$~ps. 
		(c) ((d)) Average displacement of $Q^{(i)}_{\text{R}}$ non-linearly coupled to the laser-excited $Q_{\text{IR}(A)}$($Q_{\text{IR}(B)}$) with frequency $\Omega_{\text{IR}}=6.104$~THz ( $\Omega_{\text{IR}}=6.493$~THz). In (c) $\tau=0.3$~ps and in (d)  $\tau=0.8$~ps.}
		\label{fig:qmaxir3}
	\end{center}
\end{figure}
Now we investigate numerically the non-linear dynamics of the three Raman phonon modes of interest in response to the excitation of a single IR mode. We consider the IR modes with frequencies $\Omega_{\text{IR}}=6.104$~THz ($Q_{\text{IR}(A)}$) and $\Omega_{\text{IR}}=6.493$~THz ($Q_{\text{IR}(B)}$), which couple to electric fields parallel to the $y$- and $x$-directions, respectively. The numerical solutions for $Q_{\text{IR}(A)}$ and $Q_{\text{R}(i)}$ are shown in Fig. \ref{fig:qmaxir3}(a-b) for $E_0= 4$ MV/cm, and $\tau=0.2$~ps. $Q_{\text{R}(2)}$ shows the largest amplitude, followed by $Q_{\text{R}(1)}$ which involves displacements in the $\hat z$-direction, perpendicular to the layers. $Q_{\text{R}(3)}$ with $B_g$ representation does not participate in the dynamics due to the weak coupling with $Q_{\text{R}(1)}$ and $Q_{\text{R}(2)}$, and the absence coupling with $Q_{\text{IR}(A)}$ at cubic order. In Fig. \ref{fig:qmaxir3}(c) ((d)), we plot the averaged displacements $\langle Q_{\text{R}(i)}\rangle$ as a function of $E_0$ in response to excitation of $Q_{\text{IR}(A)}$ ($\Omega_{\text{IR}(B)}$) with $\tau= 0.3$~ps ($\tau= 0.8$~ps). Therefore, the direction of the shift between the layers can be control by selectively exciting $\Omega_{\text{IR}(A)}$ or $\Omega_{\text{IR}(B)}$. Next, we will study the magnetic order and show that the induced layer displacements accessible using non-linear phonon processes can switch the sign of the interlayer exchange interactions.

%%%%%%%%%%%%%%%%%%%%%%%%%%%%%%%%%%
% Spin Hamiltonian
%%%%%%%%%%%%%%%%%%%%%%%%%%%%%%%%%%

\textit{Effective spin interaction}. Recently, theoretical work~\cite{Xu2018_cri3, xu2020} and a combined study employing group theory and ferromagnetic resonance measurements~\cite{lee2020fundamental} proposed that CrI$_3$ is described by the Heisenberg-Kitaev~\cite{kitaev2006,rau2014} Hamiltonian $\mathcal H = \mathcal H_{\text{intra}}+\mathcal H_{\text{inter}}$, where the intralayer Hamiltonian is given by
\begin{align}\nonumber
\mathcal H_{\text{intra}} &=    \sum_{ \langle ij \rangle \in \lambda \mu (\nu) } \mathcal{J} \ss_i \cdot \ss_j + K s^{\nu}_i s^{\nu}_j + \Gamma \left( s^{\lambda}_i s^{\mu}_j + s^{\mu}_i s^{\lambda}_j \right),
\label{eq:spin_ham_intra}
\end{align}
and contains $\mathcal{J}$ Heisenberg and $K$ Kitaev~\cite{kitaev2006} interactions with off-diagonal exchange $\Gamma$ ~\cite{rau2014}. The Heisenberg-Kitaev Hamiltonian $\mathcal H_{\text{intra}}$ has been studied extensively. For example, the equilibrium phase diagram~\cite{rau2014} and the magnon contribution to thermal conductivity has been determined~\cite{stamokostas2017}, and the spin-wave spectrum has been shown to carry nontrivial Chern numbers~\cite{McClarty2018}. In Ref.~[\onlinecite{lee2020fundamental}], the intra-layer interaction constants for CrI$_3$ were determined experimentally to be $\mathcal{J} = - 0.2$~meV, $K=-5.2$~meV, and $\Gamma=-67.5$~$\mu \text{eV}$.

In experiments~\cite{chen2018,lee2020fundamental}, the interlayer Hamiltonian has been assumed to be $\mathcal H_{\text{inter}}= \sum_{\langle i j \rangle \in \mbox{int.}} J_{\perp} \ss_i \cdot \ss_j$, with $|J_{\perp}|=0.03$~meV in Ref. [\onlinecite{lee2020fundamental}], and $|J_{\perp}|=0.59$~meV in Ref. [\onlinecite{chen2018}], as extracted from ferromagnetic resonance and inelastic neutron scattering measurements in bilayer and bulk CrI$_3$, respectively. Although both experiments propose different intralayer spin models, both find that the interlayer energy scale is much smaller than the intralayer energy scale. Here, we map the interlayer Hamiltonian into a Heisenberg model of the form $\mathcal H_{\text{inter}}=\frac{1}{2} \sum_{ij \in \text{int.}} J_{ij} \ss_i \cdot \ss_j$, and determine $J_{ij}$ from first principles (generalized gradient approximation with Hubbard U=$1$~eV fixed to reproduce the b-CrI$_3$ critical temperature $T_C=45$~K) using a Green's function approach and the magnetic force theorem (for a detailed explanation of the method and applications, see Ref.~[\onlinecite{hoffmann2020}]).  

The coupling between the spin and the phonons enters through the interatomic distance dependence of the exchange constants~\cite{granado1999}. Under a lattice deformation, and for small deviations from the equilibrium position, the exchange interaction is given by
%
%\begin{equation}
$
J[ \boldsymbol{u}(t) ] = J^0 +  \delta J \hat \dd \cdot   \boldsymbol{u}(t) + \mathcal O(\boldsymbol{u}(t)^2),
$
%\end{equation}
%
where $J^0$ corresponds to the equilibrium interaction, $\delta J$ is the strength of the first-order correction in the direction $\boldsymbol{\hat \delta}$, and $\boldsymbol{u}(t)$ is the real-space phonon displacement. Given that the infrared phonon frequencies we propose to use ($\Omega_{\text{IR}}\approx 6.49,6.1$~THz) are much larger than the relevant interlayer interactions ($\lesssim 1$~meV), to leading order, Floquet theory indicates that the effective interlayer exchange interaction becomes 
%\begin{equation}
$J^{\text{eff}} = J^0 +  \delta J \boldsymbol{\hat \delta} \cdot \langle \boldsymbol{u}_{\text{R}} \rangle,$
%\end{equation}
where  $\langle u_{\text{R}} \rangle$ is the time-averaged Raman mode displacement. Therefore, in order to determine the effect of the non-linear phonon displacements, we compute the effective exchange interactions in b-CrI$_3$ for layers displaced with respect to each other in the direction of the low-frequency Raman modes. 

\begin{figure}[t]
	\begin{center}
		\includegraphics[width=8.5cm]{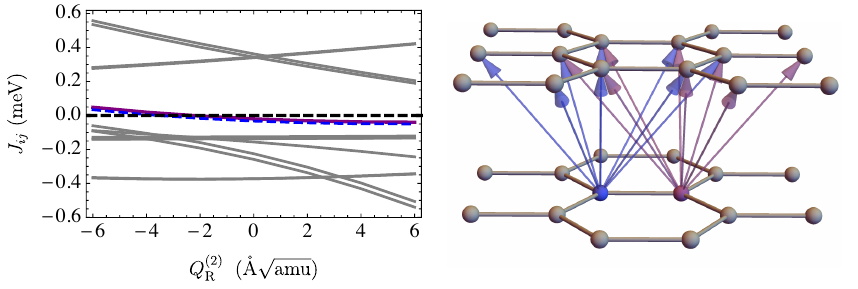}
		\caption{(Color online) Left: Interlayer exchange interaction among magnetic moments, up to third-order nearest-neighbors, as a function of the Raman displacement $Q_{\text{R}(2)}$. Right: b-CrI$_3$ sketch, where the arrows indicate the nearest and next-nearest neighbors. }
		\label{fig:js}
	\end{center}
\end{figure}

The interlayer exchange interactions $J_{ij}$ (up to third-order nearest neighbors) are shown in Fig. \ref{fig:js} as a function of the Raman displacement amplitude $Q_{\text{R}(2)}$, revealing the complexity of the interlayer magnetic order in b-CrI$_3$. In order to compare our theoretical result for the interlayer interaction with experiments, we define $J_{\perp} \equiv (1/2)\sum_{ij} J_{ij}$. The effective Floquet exchange interaction is then $J^{\text{eff}}_{\perp}(\langle Q_{\text{R}(2)} \rangle ) =J^0_{\perp} + \delta J_{\perp} \langle Q_{\text{R}(2)} \rangle$. We find $J^0_{\perp}= -0.366 $~meV and $\delta J_{\perp} =-0.0713$~meV/$(\text{\AA}\sqrt{\text{amu}})$, with $J^{\text{eff}}>0$, thus preferring FM order, for $\langle Q_{\text{R}(2)} \rangle < -5.13\text{\AA}\sqrt{\text{amu}}$ which corresponds to a real-space displacement of $\sim 3.13\%$ of the Cr-Cr interatomic distance. However, $J^0_{\perp}$ overestimates the experimental value for b-CrI$_3$~\cite{lee2020fundamental}. Using $J^0_{\perp}$ as a fitting parameter from experiments, and $\delta J_{\perp}$ from our calculations, we find $J^{\text{eff}}(\langle Q_{\text{R}(2)} \rangle)>0$ for $\langle Q_{\text{R}(2)} \rangle < -0.42\text{\AA}\sqrt{\text{amu}}$, $\sim0.3\%$ of the Cr-Cr interatomic distance.

Experimentally, $\langle Q_{\text{R}(2)} \rangle \approx 0.5$ \AA $\sqrt{\text{amu}}$ can be achieved by driving $Q_{\text{IR}(A)}$ with $E_0= 4$~MV$/$cm and $\tau=0.2$~ps. This leads to $Q_{\text{IR}(A)}$ maximum amplitude oscillations $\sim 2$ \AA $\sqrt{\text{amu}}$. 
On the other hand, driving $Q_{\text{IR}(B)}$ with $E_0=20$~MV$/$cm and $\tau=0.8$~ps, leads to $\langle Q_{\text{R}(2)} \rangle \approx - 0.53$ \AA $\sqrt{\text{amu}}$ with $Q_{\text{IR}(B)}$ reaching maximum amplitude oscillations $\sim 1.6$ \AA $\sqrt{\text{amu}}$. Notice that to obtain negative $\langle Q_{\text{R}(2)} \rangle$ displacements, stronger electric fields are required due to the relatively weak coupling of $Q_{\text{IR}(B)}$ with the laser pulse, since the effective charge is $|Z_B^*|=0.034$~$e/\sqrt{\text{amu}}$, compared to $|Z_A^*|=0.740583$~$e/\sqrt{\text{amu}}$ for $Q_{\text{IR}(A)}$.

\textit{Conclusions}. In this work, we studied theoretically b-CrI$_3$ driven with low-frequency light pulses. We found that coherently driving an infrared mode can activate low-frequency Raman modes involving relative displacements between the layers, which oscillate around new shifted equilibrium positions due to non-linear phonon processes. These relatively small transient lattice distortions can modify the exchange interactions and change the sign of the interlayer interaction. This provides the opportunity to change the magnetic order in a system via low-frequency drives.  Similar results should be possible for other layer magnetic materials with weak inter-layer bonds.

\textit{Acknowledgments}. We thank Matthias Geilhufe for useful discussions on group theory and the use of GTPack, and D. M. Juraschek for useful discussions on non-linear phononics. This research was primarily supported by the National Science Foundation through the Center for Dynamics and Control of Materials: an NSF MRSEC under Cooperative Agreement No. DMR-1720595 and NSF Grant No. DMR-1949701. A.L. acknowledges support from the funding grant: PID2019-105488GB-I00. M.R-V thanks the hospitality of Aspen Center for Physics, supported by National Science Foundation grant PHY-1607611, where parts of this work were performed. M.G.V. acknowledges the Spanish Ministerio de Ciencia e Innovacion (grant number PID2019-109905GB-C21) and support from DFG INCIEN2019-000356 from Gipuzkoako Foru Aldundia

% bibliography
%\bibliography{phonons_floquet}


\appendix 

\section{$\Gamma$-point phonon frequencies and macroscopic dielectric tensor}
\label{app:dft}

In this section we list all the phonon frequencies at the $\Gamma$ point, the macroscopic dielectric tensor, and the Born charges. We use five different approaches, which show a good agreement among all of them. Table [\ref{tab:LPer}] shows converged phonon frequencies at the $\Gamma$ point of the CrI$_3$ bilayer with the C2/m space group. The different approaches we employ are described below:

\textbf{qe1:} QUANTUM ESPRESSO \cite{Giannozzi_2009,Giannozzi_2017} calculation using Density Functional Perturbation Theory (dfpt). GGA-PAW potentials were employed with Ecut=55 Ry and Ecutrho=490 Ry. k-grid sampling of 12x12x1 and Van der Waals correction of type grimme-d2 were used. 

\textbf{qe2:} same as qe1, imposing the acoustic sum rule to the dynamical matrix, $\sum_{\mathbf{L},j}C_{\alpha i,\beta j}(\mathbf{R_L})=0$. Corrects the non analytic contribution to LO $\omega_L=\sqrt{\omega_o^2+4\pi^2Z^{*2}/\Omega\epsilon_{\infty}M}$ are also included. 

\textbf{vasp3:} VASP\cite{kresse1996,kresse1996b} calculation using finite differences (fd) (IBRION=6, NFREE=4). GGA-PAW potentials were employed with  ENCUT$=$600eV.  k-grid sampling of 12x12x1 and Van der Waals correction of type grimme-d2 were used. 

\textbf{vasp4:} VASP calculation using finite differences (IBRION=6 NFREE=4). LDA-PAW potentials were employed with  ENCUT$=$600eV.  k-grid sampling of 12x12x1 and NO VdW correction.

\textbf{vasp5:} VASP calculation using DFPT (IBRION=8). LDA-PAW potentials were employed with  ENCUT$=$600eV.  k-grid sampling of 12x12x1 and NO VdW correction.\\

\begin{table}[]
\begin{tabular}{|l|l|l|}
\hline
3.86908   & 0.000000      & -0.004342 \\ \hline
0.000000       & 3.869087 & 0.000000       \\ \hline
-0.004340 & 0.000000      & 1.490740  \\ \hline
\end{tabular}
\caption{Macroscopic dielectric tensor in cartesian coordinates as obtained with QE and  DFPT. }
\label{tab:mdtqe1}
\end{table}

\begin{table}[]
\begin{tabular}{|l|l|l|}
\hline
4.009906 & 0.000000 & -0.005896 \\ 
 \hline
 0.000000 & 4.011021 & 0.000000  \\
 \hline
 -0.005896 & 0.000000 & 1.491395  \\ \hline
\end{tabular}
\caption{Macroscopic dielectric tensor in cartesian coordinates as obtained with VASP and DFPT. }
\label{tab:mdtqe1}
\end{table}

\begin{table*}
\begin{tabular}{| c | c | c | c | c | l | c | c | c | c | c | c | c } % creating 10 columns
\hline % inserting double-line
 & \multicolumn{3}{c|}{GGA + VdW}  & \multicolumn{2}{c|}{LDA} &
 & \multicolumn{3}{c|}{GGA + VdW}  & \multicolumn{2}{c|}{LDA} \\ 
 \hline
\# & \makecell{qe1\\ dfpt} & \makecell{qe2\\ dfpt} & \makecell{vasp3\\ fd} & \makecell{vasp4\\ fd} & \makecell{vasp5\\ dfpt} &\# & \makecell{qe1\\ dfpt} & \makecell{qe2\\ dfpt} & \makecell{vasp3\\ fd} & \makecell{vasp4\\ fd} & \makecell{vasp5\\ dfpt} \\
\hline % inserts single-line
4 &  0.2385 & 0.4976 & 0.4604 & 0.3783 & $\mathbb{C}$ & 27 & 3.2697 & 3.2740 & 3.2433 & 3.2673 & 3.2269 \\
5 & 0.3690 &  0.4989 & 0.4670 & 0.3864 & 0.1714 & 28 & 3.2807 & 3.2937 & 3.2450 & 3.2891 & 3.2303\\
6 & 0.4477 &  0.9079 & 0.9595 & 0.7972 & 0.9079 &  29 & 3.4095 & 3.4022 & 3.3609 & 3.4638 & 3.4018\\
7 & 1.6958 & 1.7132 & 1.6860 & 1.5308 & 1.4793 & 30 & 3.4183 & 3.4218 & 3.3665 & 3.4722 & 3.4203\\
8 & 1.7237 & 1.7377 & 1.7040 & 1.5371 & 1.5042 & 31 & 3.4452 & 3.4397 & 3.3920 & 3.4741 & 3.4293\\
9 & 1.7367 & 1.7564 & 1.7141 & 1.5562 & 1.5239 & 32 & 3.4454 & 3.4485 & 3.3983 & 3.4783 & 3.4310\\
10 & 1.7413 & 1.7628 & 1.7288 & 1.5587 & 1.5372 & 33 & 3.8140 & 3.8112 & 3.7515 & 3.9069 & 3.9070\\
11 & 2.0703 & 2.1097 & 2.0503 & 1.7126 & 1.6859 &  34 & 3.9230 & 3.9243 & 3.8707 & 3.9697 & 3.9532\\
12 & 2.0853 & 2.1280 & 2.0574 & 1.7243 & 1.6976 & 35 & 4.0210 & 4.0156 & 3.9941 & 4.0491 & 4.0344\\
13 & 2.5990 & 2.6087 & 2.5759 & 2.2839 & 2.2798 & 36 & 4.0290 & 4.0221 & 3.9950 & 4.0494 & 4.0366\\
14 & 2.6116 & 2.6247 & 2.5833 & 2.3164 & 2.3034 & 37 & 5.5624 & 5.9087 & 5.8083 & 6.5331 & 6.5479\\
15 & 2.6443 & 2.6620 & 2.6371 & 2.4387 & 2.3883 & 38 & 5.6466 & 6.0482 & 5.8098 & 6.5354 & 6.5505\\
16 & 2.6632 & 2.6722 & 2.6418 & 2.4546 & 2.4212 &  39 & 6.0618 & 6.1663 & 6.0943 & 6.8176 & 6.8098\\
17 & 2.6633 & 2.6754 & 2.6473 & 2.4709 & 2.4273 & 40 & 6.1175 & 6.1734 & 6.1014 & 6.8193 & 6.8138\\
18 & 2.7513 & 2.7614 & 2.7048 & 2.4725 & 2.4423 & 41 & 6.1473 & 6.1793 & 6.1042 & 6.8248  & 6.8228 \\
19 & 2.7664 & 2.9146 & 2.8981 & 2.6894 & 2.7276 & 42 & 6.1858 & 6.2221 & 6.1201 & 6.8332 & 6.8248 \\
20 & 2.8047 & 2.9525 & 2.9078 & 2.6906 & 2.7291 & 43 & 6.5188 & 6.5480 & 6.4940 & 7.2969 & 7.2970 \\
21 & 3.0059 & 3.0036 & 2.9605 & 3.0827 & 3.0360 &  44 & 6.5201 & 6.5638 & 6.5038  & 7.3029 & 7.3030 \\
22 & 3.0340 & 3.0339 & 2.9907 & 3.0894 & 3.0445 & 45 & 6.5378 & 6.6344 & 6.5193 & 7.3076 & 7.3071 \\
23 & 3.0363 & 3.0376 & 3.0041 & 3.0988 & 3.0666 & 46 & 6.6435 & 6.6723 & 6.5230 & 7.3136 & 7.3132 \\
24 & 3.0384 & 3.0417 & 3.0092 & 3.1003 & 3.0679 & 47 & 7.0222 & 7.3521 & 7.2560 & 7.9413  & 7.9799 \\
25 & 3.2498 & 3.2580 & 3.2352 & 3.2557 & 3.1649 & 48 & 7.0914 & 7.4011 & 7.2676 & 7.9484 & 7.9914 \\
26 & 3.2694 & 3.2730 & 3.2365 & 3.2628 & 3.2023 & &  &  &  & & \\
% [1ex] adds vertical space
\hline
% inserts single-line
\end{tabular}
\caption{\label{tab:LPer} $\Gamma$-point phonon frequencies for bilayer CrI$_3$ with space group C2/m. The method abbreviations are: qe $=$ Quantum-Espresso, dfpt$=$ Density Functional Perturbation Theory, fd$=$ Finite difference}
\end{table*}

\subsection{Born effective charges}
The effective charge tensors $Z^*_{\kappa, ij} $ (units of the electron electric charge $e$ and in cartesian axis) of atom $\kappa$ are listed below. The Born effective charge is defined as \cite{gonze1997, baroni2001}
\begin{equation}
Z^*_{\alpha, i} = \sum_{\kappa,j} Z^*_{\kappa, ij} \frac{e_{\alpha,\kappa,j}}{\sqrt{m_\kappa}},
\end{equation}
where $\alpha$ labels the phonon mode, $i$ the direction in cartesian coordinates,  $\kappa$ labels the atoms in the unit cell, $m_\kappa$ is the mass of atom $\kappa$, and $e_{\alpha,\kappa,j}$ corresponds to the dynamical matrix eigenvector $\alpha$, atom $\kappa$, in the direction $j$ normalized as $\sum_{\kappa,j} (e_{\alpha,\kappa,j})^*e_{\beta,\kappa,j} = \delta_{\alpha \beta}$.

\begin{table}[H]
\begin{tabular}{lllllll}
\multicolumn{3}{c}{atom 1 Cr}                                                                  &                       & \multicolumn{3}{c}{atom 9 I}                                                                  \\ \cline{1-3} \cline{5-7} 
\multicolumn{1}{|l|}{2.50670}  & \multicolumn{1}{l|}{0.01211}  & \multicolumn{1}{l|}{0.0035}   & \multicolumn{1}{l|}{} & \multicolumn{1}{l|}{0.06017}  & \multicolumn{1}{l|}{0.09685}  & \multicolumn{1}{l|}{-0.10598} \\ \cline{1-3} \cline{5-7} 
\multicolumn{1}{|l|}{0.01266}  & \multicolumn{1}{l|}{2.48423}  & \multicolumn{1}{l|}{-0.00869} & \multicolumn{1}{l|}{} & \multicolumn{1}{l|}{-0.41963} & \multicolumn{1}{l|}{-0.59530} & \multicolumn{1}{l|}{-0.46809} \\ \cline{1-3} \cline{5-7} 
\multicolumn{1}{|l|}{-0.00018} & \multicolumn{1}{l|}{-0.00217} & \multicolumn{1}{l|}{0.27999}  & \multicolumn{1}{l|}{} & \multicolumn{1}{l|}{0.06017}  & \multicolumn{1}{l|}{-0.09685} & \multicolumn{1}{l|}{-0.10598} \\ \cline{1-3} \cline{5-7} 
\multicolumn{3}{c}{atom 2 Cr}                                                                  &                       & \multicolumn{3}{c}{atom 10 I}                                                                 \\ \cline{1-3} \cline{5-7} 
\multicolumn{1}{|l|}{2.50670}  & \multicolumn{1}{l|}{-0.01211} & \multicolumn{1}{l|}{0.00352}  & \multicolumn{1}{l|}{} & \multicolumn{1}{l|}{-1.10547} & \multicolumn{1}{l|}{0.43882}  & \multicolumn{1}{l|}{0.27730}  \\ \cline{1-3} \cline{5-7} 
\multicolumn{1}{|l|}{-0.01266} & \multicolumn{1}{l|}{2.48423}  & \multicolumn{1}{l|}{0.00869}  & \multicolumn{1}{l|}{} & \multicolumn{1}{l|}{0.43112}  & \multicolumn{1}{l|}{-0.55676} & \multicolumn{1}{l|}{0.44096}  \\ \cline{1-3} \cline{5-7} 
\multicolumn{1}{|l|}{-0.00018} & \multicolumn{1}{l|}{0.00217}  & \multicolumn{1}{l|}{0.27999}  & \multicolumn{1}{l|}{} & \multicolumn{1}{l|}{0.05116}  & \multicolumn{1}{l|}{0.08974}  & \multicolumn{1}{l|}{-0.0775}  \\ \cline{1-3} \cline{5-7} 
\multicolumn{3}{c}{atom 3 Cr}                                                                  &                       & \multicolumn{3}{c}{atom 11 I}                                                                 \\ \cline{1-3} \cline{5-7} 
\multicolumn{1}{|l|}{2.50670}  & \multicolumn{1}{l|}{-0.01211} & \multicolumn{1}{l|}{0.00352}  & \multicolumn{1}{l|}{} & \multicolumn{1}{l|}{-0.37557} & \multicolumn{1}{l|}{0.00000}  & \multicolumn{1}{l|}{-0.55006} \\ \cline{1-3} \cline{5-7} 
\multicolumn{1}{|l|}{-0.01266} & \multicolumn{1}{l|}{2.48423}  & \multicolumn{1}{l|}{0.00869}  & \multicolumn{1}{l|}{} & \multicolumn{1}{l|}{0.00000}  & \multicolumn{1}{l|}{-1.32030} & \multicolumn{1}{l|}{0.00000}  \\ \cline{1-3} \cline{5-7} 
\multicolumn{1}{|l|}{-0.00018} & \multicolumn{1}{l|}{0.00218}  & \multicolumn{1}{l|}{0.28094}  & \multicolumn{1}{l|}{} & \multicolumn{1}{l|}{-0.11557} & \multicolumn{1}{l|}{0.00000}  & \multicolumn{1}{l|}{-0.10817} \\ \cline{1-3} \cline{5-7} 
\multicolumn{3}{c}{atom 4 Cr}                                                                  &                       & \multicolumn{3}{c}{atom 12 I}                                                                 \\ \cline{1-3} \cline{5-7} 
\multicolumn{1}{|l|}{2.50670}  & \multicolumn{1}{l|}{0.01211}  & \multicolumn{1}{l|}{0.00352}  & \multicolumn{1}{l|}{} & \multicolumn{1}{l|}{-0.31861} & \multicolumn{1}{l|}{0.00000}  & \multicolumn{1}{l|}{-0.55899} \\ \cline{1-3} \cline{5-7} 
\multicolumn{1}{|l|}{0.01266}  & \multicolumn{1}{l|}{2.48423}  & \multicolumn{1}{l|}{-0.00869} & \multicolumn{1}{l|}{} & \multicolumn{1}{l|}{0.00000}  & \multicolumn{1}{l|}{-1.37494} & \multicolumn{1}{l|}{0.00000}  \\ \cline{1-3} \cline{5-7} 
\multicolumn{1}{|l|}{-0.00018} & \multicolumn{1}{l|}{-0.00218} & \multicolumn{1}{l|}{0.28094}  & \multicolumn{1}{l|}{} & \multicolumn{1}{l|}{-0.10654} & \multicolumn{1}{l|}{0.00000}  & \multicolumn{1}{l|}{-0.11383} \\ \cline{1-3} \cline{5-7} 
\multicolumn{3}{c}{atom 5 I}                                                                   &                       & \multicolumn{3}{c}{atom 13 I}                                                                 \\ \cline{1-3} \cline{5-7} 
\multicolumn{1}{|l|}{-0.31877} & \multicolumn{1}{l|}{0.00000}  & \multicolumn{1}{l|}{-0.55912} & \multicolumn{1}{l|}{} & \multicolumn{1}{l|}{-1.06967} & \multicolumn{1}{l|}{-0.41268} & \multicolumn{1}{l|}{0.27350}  \\ \cline{1-3} \cline{5-7} 
\multicolumn{1}{|l|}{0.00000}  & \multicolumn{1}{l|}{-1.37542} & \multicolumn{1}{l|}{0.00000}  & \multicolumn{1}{l|}{} & \multicolumn{1}{l|}{-0.41970} & \multicolumn{1}{l|}{-0.59544} & \multicolumn{1}{l|}{-0.46806} \\ \cline{1-3} \cline{5-7} 
\multicolumn{1}{|l|}{-0.10597} & \multicolumn{1}{l|}{0.00000}  & \multicolumn{1}{l|}{-0.11291} & \multicolumn{1}{l|}{} & \multicolumn{1}{l|}{0.05985}  & \multicolumn{1}{l|}{-0.09629} & \multicolumn{1}{l|}{-0.10442} \\ \cline{1-3} \cline{5-7} 
\multicolumn{3}{c}{atom 6  I}                                                                  &                       & \multicolumn{3}{c}{atom 14 I}                                                                 \\ \cline{1-3} \cline{5-7} 
\multicolumn{1}{|l|}{-0.37544} & \multicolumn{1}{l|}{0.00000}  & \multicolumn{1}{l|}{-0.55003} & \multicolumn{1}{l|}{} & \multicolumn{1}{l|}{-1.10520} & \multicolumn{1}{l|}{0.43869}  & \multicolumn{1}{l|}{0.27732}  \\ \cline{1-3} \cline{5-7} 
\multicolumn{1}{|l|}{0.00000}  & \multicolumn{1}{l|}{-1.32046} & \multicolumn{1}{l|}{0.00000}  & \multicolumn{1}{l|}{} & \multicolumn{1}{l|}{0.43116}  & \multicolumn{1}{l|}{-0.55665} & \multicolumn{1}{l|}{0.44089}  \\ \cline{1-3} \cline{5-7} 
\multicolumn{1}{|l|}{-0.11621} & \multicolumn{1}{l|}{0.00000}  & \multicolumn{1}{l|}{-0.1097}  & \multicolumn{1}{l|}{} & \multicolumn{1}{l|}{0.05145}  & \multicolumn{1}{l|}{0.09024}  & \multicolumn{1}{l|}{-0.07844} \\ \cline{1-3} \cline{5-7} 
\multicolumn{3}{c}{atom 7 I}                                                                   &                       & \multicolumn{3}{c}{atom 15 I}                                                                 \\ \cline{1-3} \cline{5-7} 
\multicolumn{1}{|l|}{-1.10547} & \multicolumn{1}{l|}{-0.43882} & \multicolumn{1}{l|}{0.27730}  & \multicolumn{1}{l|}{} & \multicolumn{1}{l|}{-1.10520} & \multicolumn{1}{l|}{-0.43869} & \multicolumn{1}{l|}{0.27732}  \\ \cline{1-3} \cline{5-7} 
\multicolumn{1}{|l|}{-0.43112} & \multicolumn{1}{l|}{-0.55676} & \multicolumn{1}{l|}{-0.44096} & \multicolumn{1}{l|}{} & \multicolumn{1}{l|}{-0.43116} & \multicolumn{1}{l|}{-0.55665} & \multicolumn{1}{l|}{-0.44089} \\ \cline{1-3} \cline{5-7} 
\multicolumn{1}{|l|}{0.05116}  & \multicolumn{1}{l|}{-0.08974} & \multicolumn{1}{l|}{-0.07759} & \multicolumn{1}{l|}{} & \multicolumn{1}{l|}{0.05145}  & \multicolumn{1}{l|}{-0.09024} & \multicolumn{1}{l|}{-0.07844} \\ \cline{1-3} \cline{5-7} 
\multicolumn{3}{c}{atom 8 I}                                                                   &                       & \multicolumn{3}{c}{atom 16 I}                                                                 \\ \cline{1-3} \cline{5-7} 
\multicolumn{1}{|l|}{-1.06968} & \multicolumn{1}{l|}{0.41271}  & \multicolumn{1}{l|}{0.27359}  & \multicolumn{1}{l|}{} & \multicolumn{1}{l|}{-1.06967} & \multicolumn{1}{l|}{0.41268}  & \multicolumn{1}{l|}{0.27350}  \\ \cline{1-3} \cline{5-7} 
\multicolumn{1}{|l|}{0.41963}  & \multicolumn{1}{l|}{-0.59530} & \multicolumn{1}{l|}{0.46809}  & \multicolumn{1}{l|}{} & \multicolumn{1}{l|}{0.41970}  & \multicolumn{1}{l|}{-0.59544} & \multicolumn{1}{l|}{0.46806}  \\ \cline{1-3} \cline{5-7} 
\multicolumn{1}{|l|}{0.06017}  & \multicolumn{1}{l|}{0.09685}  & \multicolumn{1}{l|}{-0.10598} & \multicolumn{1}{l|}{} & \multicolumn{1}{l|}{0.05985}  & \multicolumn{1}{l|}{0.09629}  & \multicolumn{1}{l|}{-0.10442} \\ \cline{1-3} \cline{5-7} 
\end{tabular}
\caption{Effective charge tensors $Z^*_{\kappa, ij} $ (units of the electron electric charge $e$ and in cartesian axis) for each atom $\kappa$ in the unit cell obtained with the \textbf{qe1} method.}
\label{tab:qe1}
\end{table}
\begin{table}[H]
\begin{tabular}{lllllll}
\multicolumn{3}{c}{atom 1 Cr}                                                                  &                       & \multicolumn{3}{c}{atom 9 I}                                                                  \\ \cline{1-3} \cline{5-7} 
\multicolumn{1}{|l|}{2.47034}  & \multicolumn{1}{l|}{0.00975}  & \multicolumn{1}{l|}{0.00028}  & \multicolumn{1}{l|}{} & \multicolumn{1}{l|}{-1.05146} & \multicolumn{1}{l|}{-0.41069} & \multicolumn{1}{l|}{0.28647}  \\ \cline{1-3} \cline{5-7} 
\multicolumn{1}{|l|}{0.01476}  & \multicolumn{1}{l|}{2.45196}  & \multicolumn{1}{l|}{-0.01247} & \multicolumn{1}{l|}{} & \multicolumn{1}{l|}{-0.42081} & \multicolumn{1}{l|}{-0.57647} & \multicolumn{1}{l|}{-0.48663} \\ \cline{1-3} \cline{5-7} 
\multicolumn{1}{|l|}{-0.00207} & \multicolumn{1}{l|}{-0.00222} & \multicolumn{1}{l|}{0.27215}  & \multicolumn{1}{l|}{} & \multicolumn{1}{l|}{0.05816}  & \multicolumn{1}{l|}{-0.09177} & \multicolumn{1}{l|}{-0.09762} \\ \cline{1-3} \cline{5-7} 
\multicolumn{3}{c}{atom 2 Cr}                                                                  &                       & \multicolumn{3}{c}{atom 10 I}                                                                 \\ \cline{1-3} \cline{5-7} 
\multicolumn{1}{|l|}{2.47034}  & \multicolumn{1}{l|}{-0.00975} & \multicolumn{1}{l|}{0.00028}  & \multicolumn{1}{l|}{} & \multicolumn{1}{l|}{-1.08708} & \multicolumn{1}{l|}{0.43826}  & \multicolumn{1}{l|}{0.28929}  \\ \cline{1-3} \cline{5-7} 
\multicolumn{1}{|l|}{-0.01475} & \multicolumn{1}{l|}{2.45262}  & \multicolumn{1}{l|}{0.01258}  & \multicolumn{1}{l|}{} & \multicolumn{1}{l|}{0.43034}  & \multicolumn{1}{l|}{-0.53772} & \multicolumn{1}{l|}{0.45867}  \\ \cline{1-3} \cline{5-7} 
\multicolumn{1}{|l|}{-0.00207} & \multicolumn{1}{l|}{0.00222}  & \multicolumn{1}{l|}{0.27215}  & \multicolumn{1}{l|}{} & \multicolumn{1}{l|}{0.05298}  & \multicolumn{1}{l|}{0.09212}  & \multicolumn{1}{l|}{-0.07155} \\ \cline{1-3} \cline{5-7} 
\multicolumn{3}{c}{atom 3 Cr}                                                                  &                       & \multicolumn{3}{c}{atom 11 I}                                                                 \\ \cline{1-3} \cline{5-7} 
\multicolumn{1}{|l|}{2.47028}  & \multicolumn{1}{l|}{-0.00955} & \multicolumn{1}{l|}{0.00023}  & \multicolumn{1}{l|}{} & \multicolumn{1}{l|}{-0.36021} & \multicolumn{1}{l|}{0.00000}  & \multicolumn{1}{l|}{-0.57005} \\ \cline{1-3} \cline{5-7} 
\multicolumn{1}{|l|}{-0.01476} & \multicolumn{1}{l|}{2.45196}  & \multicolumn{1}{l|}{0.01247}  & \multicolumn{1}{l|}{} & \multicolumn{1}{l|}{-0.00050} & \multicolumn{1}{l|}{-1.30942} & \multicolumn{1}{l|}{-0.00044} \\ \cline{1-3} \cline{5-7} 
\multicolumn{1}{|l|}{-0.00169} & \multicolumn{1}{l|}{0.00193}  & \multicolumn{1}{l|}{0.27170}  & \multicolumn{1}{l|}{} & \multicolumn{1}{l|}{-0.11277} & \multicolumn{1}{l|}{0.00000}  & \multicolumn{1}{l|}{-0.10142} \\ \cline{1-3} \cline{5-7} 
\multicolumn{3}{c}{atom 4 Cr}                                                                  &                       & \multicolumn{3}{c}{atom 12 I}                                                                 \\ \cline{1-3} \cline{5-7} 
\multicolumn{1}{|l|}{2.47028}  & \multicolumn{1}{l|}{0.00955}  & \multicolumn{1}{l|}{0.00023}  & \multicolumn{1}{l|}{} & \multicolumn{1}{l|}{-0.30334} & \multicolumn{1}{l|}{0.00000}  & \multicolumn{1}{l|}{-0.58193} \\ \cline{1-3} \cline{5-7} 
\multicolumn{1}{|l|}{0.01475}  & \multicolumn{1}{l|}{2.45262}  & \multicolumn{1}{l|}{-0.01258} & \multicolumn{1}{l|}{} & \multicolumn{1}{l|}{0.00061}  & \multicolumn{1}{l|}{-1.36813} & \multicolumn{1}{l|}{0.00051}  \\ \cline{1-3} \cline{5-7} 
\multicolumn{1}{|l|}{-0.00169} & \multicolumn{1}{l|}{-0.00193} & \multicolumn{1}{l|}{0.27170}  & \multicolumn{1}{l|}{} & \multicolumn{1}{l|}{-0.10721} & \multicolumn{1}{l|}{0.00000}  & \multicolumn{1}{l|}{-0.10519} \\ \cline{1-3} \cline{5-7} 
\multicolumn{3}{c}{atom 5 I}                                                                   &                       & \multicolumn{3}{c}{atom 13 I}                                                                 \\ \cline{1-3} \cline{5-7} 
\multicolumn{1}{|l|}{-0.30351} & \multicolumn{1}{l|}{0.00000}  & \multicolumn{1}{l|}{-0.58214} & \multicolumn{1}{l|}{} & \multicolumn{1}{l|}{-1.05127} & \multicolumn{1}{l|}{-0.41066} & \multicolumn{1}{l|}{0.28650}  \\ \cline{1-3} \cline{5-7} 
\multicolumn{1}{|l|}{-0.00061} & \multicolumn{1}{l|}{-1.36813} & \multicolumn{1}{l|}{-0.00051} & \multicolumn{1}{l|}{} & \multicolumn{1}{l|}{-0.42068} & \multicolumn{1}{l|}{-0.57582} & \multicolumn{1}{l|}{-0.48662} \\ \cline{1-3} \cline{5-7} 
\multicolumn{1}{|l|}{-0.10894} & \multicolumn{1}{l|}{0.00000}  & \multicolumn{1}{l|}{-0.10684} & \multicolumn{1}{l|}{} & \multicolumn{1}{l|}{0.05856}  & \multicolumn{1}{l|}{-0.09302} & \multicolumn{1}{l|}{-0.09877} \\ \cline{1-3} \cline{5-7} 
\multicolumn{3}{c}{atom 6 I}                                                                   &                       & \multicolumn{3}{c}{atom 14 I}                                                                 \\ \cline{1-3} \cline{5-7} 
\multicolumn{1}{|l|}{-0.36020} & \multicolumn{1}{l|}{0.00000}  & \multicolumn{1}{l|}{-0.56990} & \multicolumn{1}{l|}{} & \multicolumn{1}{l|}{-1.08718} & \multicolumn{1}{l|}{0.43838}  & \multicolumn{1}{l|}{0.28924}  \\ \cline{1-3} \cline{5-7} 
\multicolumn{1}{|l|}{0.00050}  & \multicolumn{1}{l|}{-1.30942} & \multicolumn{1}{l|}{0.00044}  & \multicolumn{1}{l|}{} & \multicolumn{1}{l|}{0.43028}  & \multicolumn{1}{l|}{-0.53701} & \multicolumn{1}{l|}{0.45876}  \\ \cline{1-3} \cline{5-7} 
\multicolumn{1}{|l|}{-0.10946} & \multicolumn{1}{l|}{0.00000}  & \multicolumn{1}{l|}{-0.10026} & \multicolumn{1}{l|}{} & \multicolumn{1}{l|}{0.05325}  & \multicolumn{1}{l|}{0.09106}  & \multicolumn{1}{l|}{-0.06906} \\ \cline{1-3} \cline{5-7} 
\multicolumn{3}{c}{atom 7 I}                                                                   &                       & \multicolumn{3}{c}{atom 15 I}                                                                 \\ \cline{1-3} \cline{5-7} 
\multicolumn{1}{|l|}{-1.08708} & \multicolumn{1}{l|}{-0.43826} & \multicolumn{1}{l|}{0.28929}  & \multicolumn{1}{l|}{} & \multicolumn{1}{l|}{-1.08718} & \multicolumn{1}{l|}{-0.43838} & \multicolumn{1}{l|}{0.28924}  \\ \cline{1-3} \cline{5-7} 
\multicolumn{1}{|l|}{-0.43028} & \multicolumn{1}{l|}{-0.53701} & \multicolumn{1}{l|}{-0.45876} & \multicolumn{1}{l|}{} & \multicolumn{1}{l|}{-0.43034} & \multicolumn{1}{l|}{-0.53772} & \multicolumn{1}{l|}{-0.45867} \\ \cline{1-3} \cline{5-7} 
\multicolumn{1}{|l|}{0.05298}  & \multicolumn{1}{l|}{-0.09212} & \multicolumn{1}{l|}{-0.07155} & \multicolumn{1}{l|}{} & \multicolumn{1}{l|}{0.05325}  & \multicolumn{1}{l|}{-0.09106} & \multicolumn{1}{l|}{-0.06906} \\ \cline{1-3} \cline{5-7} 
\multicolumn{3}{c}{atom 8 I}                                                                   &                       & \multicolumn{3}{c}{atom 16 I}                                                                 \\ \cline{1-3} \cline{5-7} 
\multicolumn{1}{|l|}{-1.05146} & \multicolumn{1}{l|}{0.41069}  & \multicolumn{1}{l|}{0.28647}  & \multicolumn{1}{l|}{} & \multicolumn{1}{l|}{-1.05127} & \multicolumn{1}{l|}{0.41066}  & \multicolumn{1}{l|}{0.28650}  \\ \cline{1-3} \cline{5-7} 
\multicolumn{1}{|l|}{0.42068}  & \multicolumn{1}{l|}{-0.57582} & \multicolumn{1}{l|}{0.48662}  & \multicolumn{1}{l|}{} & \multicolumn{1}{l|}{0.42081}  & \multicolumn{1}{l|}{-0.57647} & \multicolumn{1}{l|}{0.48663}  \\ \cline{1-3} \cline{5-7} 
\multicolumn{1}{|l|}{0.05816}  & \multicolumn{1}{l|}{0.09177}  & \multicolumn{1}{l|}{-0.09762} & \multicolumn{1}{l|}{} & \multicolumn{1}{l|}{0.05856}  & \multicolumn{1}{l|}{0.09302}  & \multicolumn{1}{l|}{-0.09877} \\ \cline{1-3} \cline{5-7} 
\end{tabular}
\caption{Effective charge tensors $Z^*_{\kappa, ij} $ (units of the electron electric charge $e$ and in cartesian axis) for each atom $\kappa$ in the unit cell obtained with the \textbf{vasp3} method.}
\label{tab:vasp3_z}
\end{table}

\section{Non-linear coefficients}
\label{app:nonlinear}
In this section, we calculate the non-linear coefficients for the energy potential
\begin{align}\nonumber
&V[Q_{\text{IR}},Q_{\text{R}(i)}] =  \frac{1}{2}\Omega^2_{\text{IR}} Q_{\text{IR}}^2+ \sum_{i=1}^3 \frac{1}{2}\Omega^2_{\text{R}(i)} Q^{2}_{\text{R}(i)}\\\nonumber & 
+\sum_{i=1}^2 \frac{ \beta_i}{3} Q_{\text{R(i)}}^3 + Q_{\text{IR}}^2 \sum^2_{i=1} \gamma_i Q_{\text{R(i)}} + \delta Q_{\text{R(1)}}^2 Q_{\text{R(2)}} \\ &+ \epsilon Q_{\text{R(1)}} Q_{\text{R(2)}}^2+   Q_{\text{R(3)}}^2  \sum^2_{i=1} \zeta_i Q_{\text{R(i)}} .
\label{eq:non-linear-pot}
\end{align}
shown in the main text and repeated here for reference. To determine this coefficients, we follow the procedure described in Ref.[\onlinecite{subedi2014}]. The displacement of atom $\kappa$ in the unit cell, direction $j$, in terms of the normal mode amplitude $Q_\alpha$, is given by
\begin{align}
u_{\alpha, \kappa, j } =  \frac{Q_{\alpha}}{\sqrt{m_\kappa}} e_{\alpha,\kappa,j},
\end{align} 
where $m_\kappa$ is the mass of atom $\kappa$, and $e_{\alpha,\kappa,j}$ is a dynamical matrix eigenvectors normalized as $\sum_{\kappa,j} (e_{\alpha,\kappa,j})^*e_{\beta,\kappa,j} = \delta_{\alpha \beta}$. The coefficients are listed in Table \ref{tab:non-linear-coeff}. Additionally, we considered the IR modes number 38, 31, and 29 in Table \ref{tab:LPer}, which have a larger Born effective charge than mode 43, but the corresponding coupling coefficient is negative, leading to a positive rectification. 

\begin{table}[H]
\begin{tabular}{|c|c|c|}
\hline
Coefficient & Value                  & Units                                            \\ \hline
%$\Omega^2_{\text{IR}}$    & $00$  & eV$/(\mathring{\text{A}}\sqrt{\text{amu}} )^{2}$ \\ \hline
$\Omega^{(1)^2}_{\text{R}}$    & $3.766\times 10^{-3}$  & eV$/(\mathring{\text{A}}\sqrt{\text{amu}} )^{2}$ \\ \hline
$\Omega^{(2)^2}_{\text{R}}$    & $8.64113  \times 10^{-4}$  & eV$/(\mathring{\text{A}}\sqrt{\text{amu}} )^{2}$ \\ \hline
$\Omega^{(3)^2}_{\text{R}}$    & $8.82608\times10^{-4}$  & eV$/(\mathring{\text{A}}\sqrt{\text{amu}} )^{2}$ \\ \hline
$\beta_1$   & $-2.72 \times 10^{-4}$ &eV$/(\mathring{\text{A}}\sqrt{\text{amu}} )^{3}$                                                  \\ \hline
$\beta_2$   & $2.22 \times 10^{-5}$  &  eV$/(\mathring{\text{A}}\sqrt{\text{amu}} )^{3}$                                                \\ \hline
$\gamma_1(47)$  & $-1.38 \times 10^{-4}$  & eV$/(\mathring{\text{A}}\sqrt{\text{amu}} )^{3}$                                                 \\ \hline
$\gamma_2(47)$  & $-2.83 \times 10^{-4}$ &   eV$/(\mathring{\text{A}}\sqrt{\text{amu}} )^{3}$                                               \\ \hline
$\gamma_1(41) (A)$  & $-1.56 \times 10^{-4}$  & eV$/(\mathring{\text{A}}\sqrt{\text{amu}} )^{3}$                                                 \\ \hline
$\gamma_2(41) (A)$  & $-2.66 \times 10^{-4}$ &   eV$/(\mathring{\text{A}}\sqrt{\text{amu}} )^{3}$                                               \\ \hline
$\gamma_1(43) (B)$  & $-3.15 \times 10^{-4}$  & eV$/(\mathring{\text{A}}\sqrt{\text{amu}} )^{3}$                                                 \\ \hline
$\gamma_2(43) (B)$  & $3.44 \times 10^{-4}$ &   eV$/(\mathring{\text{A}}\sqrt{\text{amu}} )^{3}$                                               \\ \hline
$\delta$    & $-7.79 \times 10^{-6}$  & eV$/(\mathring{\text{A}}\sqrt{\text{amu}} )^{3}$                                                 \\ \hline
$\epsilon$  & $-3.06 \times 10^{-5}$ &  eV$/(\mathring{\text{A}}\sqrt{\text{amu}} )^{3}$                                                \\ \hline
$\zeta_1$   & $-6.19 \times 10^{-5}$ &  eV$/(\mathring{\text{A}}\sqrt{\text{amu}} )^{3}$                                                \\ \hline
$\zeta_2$   & $-3.13 \times 10^{-5}$ &   eV$/(\mathring{\text{A}}\sqrt{\text{amu}} )^{3}$                                               \\ \hline
\end{tabular}
\caption{Non-linear coefficients obtained by fitting the energy surfaces shown in Fig. \ref{fig:non-linear}. The frequencies $\Omega^{(i)}_{\text{R}}$ are in good agreement with dynamical matrix diagonalization results. The $\gamma_i$ coefficients depend on the specific driven IR mode, and the number corresponds to Table \ref{tab:LPer}. The $\gamma_i$ coefficients labeled as A and B correspond to infrared modes $Q_{\text{IR(A)}}$ and $Q_{\text{IR(B)}}$ in the main text respectively.}
\label{tab:non-linear-coeff}
\end{table}
\begin{figure}[t]
	\begin{center}
		\includegraphics[width=8.5cm]{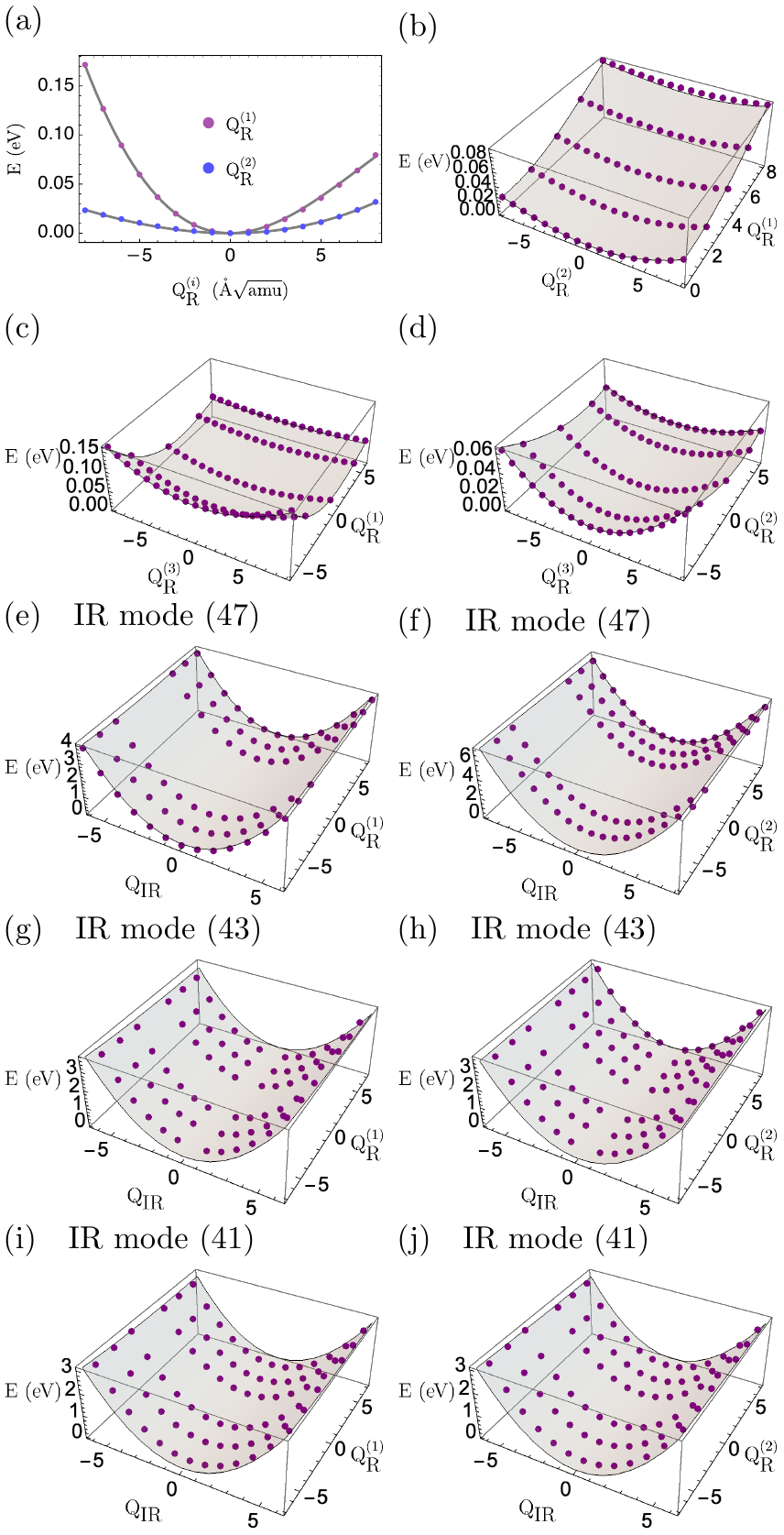}
		\caption{(Color online) Total energy as a function of the phonon mode amplitudes used to obtain the non-linear coefficents. In each panel, we vary the indicated phonon amplitudes to obtain the energy surface, keeping the other modes at equilibrium. From fitting to these curves, we can extract all the coefficients appearing in the non-linear potential Eq. (\ref{eq:non-linear-pot}). }
		\label{fig:non-linear}
	\end{center}
\end{figure}

\section{Interlayer exchange interactions.}

In this section, we provide more details on our estimates for the critical phonon amplitudes $Q_{\text{R}(i)}$, and show additional results for the interlayer exchange interactions as a function of Raman displacement $Q_{\text{R}(1)}$. 

\begin{figure}
	\begin{center}
		\includegraphics[width=8.5cm]{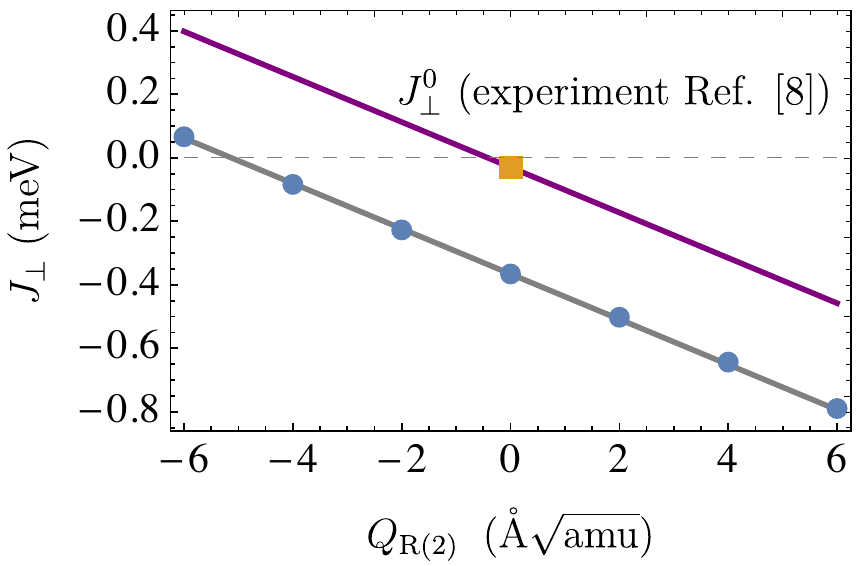}
		\caption{(Color online) Effective interlayer exchange interactions as a function of Raman displacement $Q_{\text{R}(2)}$. The blue dots correspons to our first principles calculations. The yellow square marks the experimental value for $J^0_{\perp} $.}
		\label{fig:js0}
	\end{center}
\end{figure}

In Fig. \ref{fig:js0}, we plot the effective interlayer exchange interaction defined in the main text as $J_{\perp} \equiv (1/2)\sum_{ij} J_{ij}$ in order to compare with the experimental observations. The blue dots correspond to the first principles calculations, the gray line to the fit  $J^{\text{eff}}_{\perp}(\langle Q_{\text{R}(2)} \rangle ) =J^0_{\perp} + \delta J_{\perp} \langle Q_{\text{R}(2)} \rangle$. We find $J^0_{\perp}= -0.366 $~meV and $\delta J_{\perp} =-0.0713$~meV/$(\text{\AA}\sqrt{\text{amu}})$, with $J^{\text{eff}}>0$. The yellow square corresponds the experimental measurement~\cite{lee2020fundamental}. Using this point to fit the exchange interaction at the equilibrium position, and the value extracted from the first principle calculation for $\delta J_{\perp} $, we obtain the purple curve, which we employ to estimate the transition in the most optimistic scenario and using the information available from experiments. 

\begin{figure}
	\begin{center}
		\includegraphics[width=8.5cm]{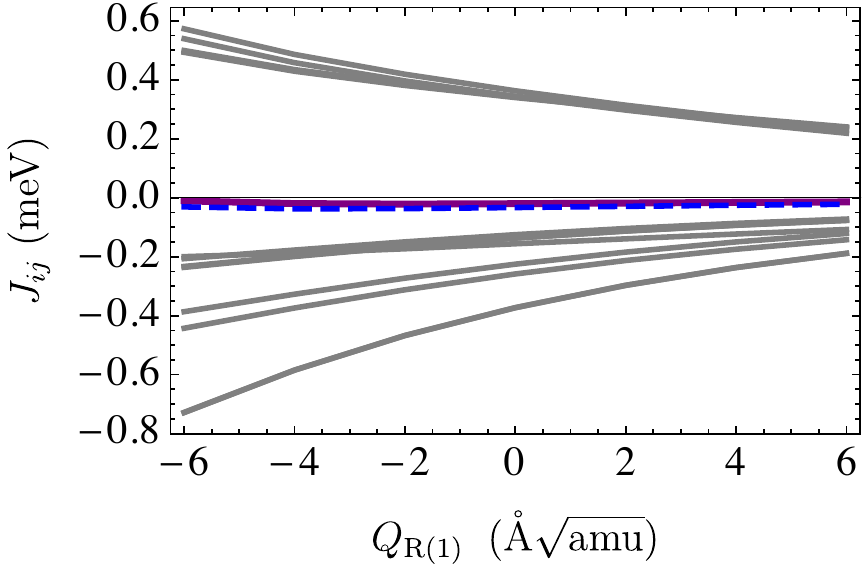}
		\caption{(Color online) Interlayer exchange interactions as a function of Raman displacement $Q_{\text{R}(1)}$.}
		\label{fig:js3}
	\end{center}
\end{figure}

\begin{figure}
	\begin{center}
		\includegraphics[width=8.5cm]{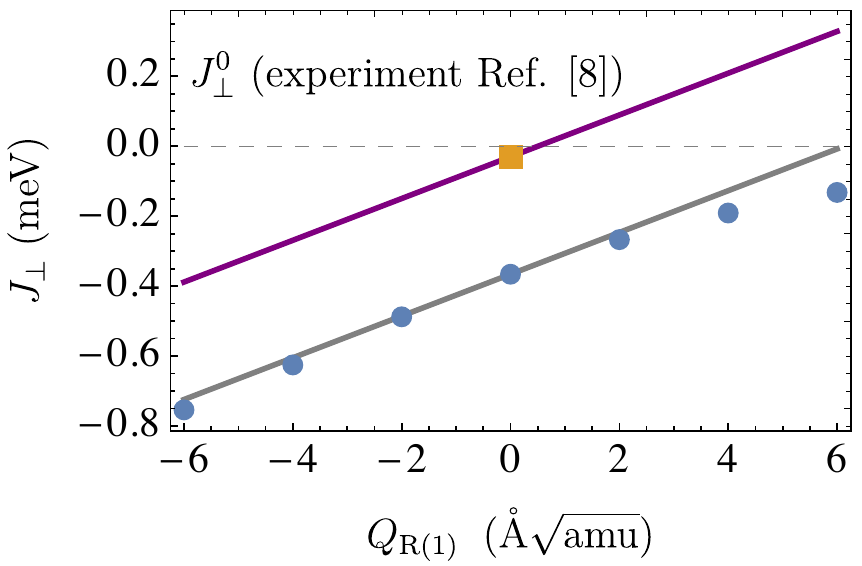}
		\caption{(Color online) Effective interlayer exchange interactions as a function of Raman displacement $Q_{\text{R}(1)}$. The blue dots correspons to our first principles calculations. The yellow square marks the experimental value for $J^0_{\perp} $.  }
		\label{fig:js4}
	\end{center}
\end{figure}

Now, we repite the same analysis for mode $Q_{\text{R}(1)}$. In Fig. \ref{fig:js3} we plot the exchange interactions per moments as a function of the Raman displacement $Q_{\text{R}(1)}$. In this case, non of the moments the individual interactions cross zero. However, the exchange interactions decrease in absolute value. In Fig. \ref{fig:js4}, we plot the effective exchange interaction. In Fig. \ref{fig:js3} , the first principles calculations (blue dots) reveal that non-linear spin-phonon couplings become relevant for  $Q_{\text{R}(1)} > 2$. Furthermore, this mode does not lead to a AFM to FM transition without taking into account input from experiments. In Fig. \ref{fig:js4}, the purple curve corresponds to $J^{\text{eff}}_{\perp}(\langle Q_{\text{R}(1)} \rangle ) =J^0_{\perp} + \delta J_{\perp} \langle Q_{\text{R}(1)} \rangle$, where the experimental $J^0_{\perp} $ has been used along with $\delta J_{\perp} = 0.0597$~meV/$(\text{\AA}\sqrt{\text{amu}})$ obtained by fitting the blue circles in the linear regime. In experiments, a displacement $\langle Q_{\text{R}(1)} \rangle = 0.5$ $\text{\AA}\sqrt{\text{amu}}$ can be achieved by driving $Q_{\text{IR}(A)}$ with $E_0= 9.5$~MV$/$cm and $\tau=0.2$~ps, which oscillated with maximum aplitudes $\{Q_{\text{IR}(A)}(t) \}_{\text{max}} \approx 4.5$ $\text{\AA}\sqrt{\text{amu}}$. Simultaneously, we have $\langle Q_{\text{R}(2)} \rangle = 2.9$ $\text{\AA}\sqrt{\text{amu}}$. 

\section{Monolayer CrI$_3$ group theory analysis}

\begin{figure}[h]
	\begin{center}
		\includegraphics[width=5.0cm]{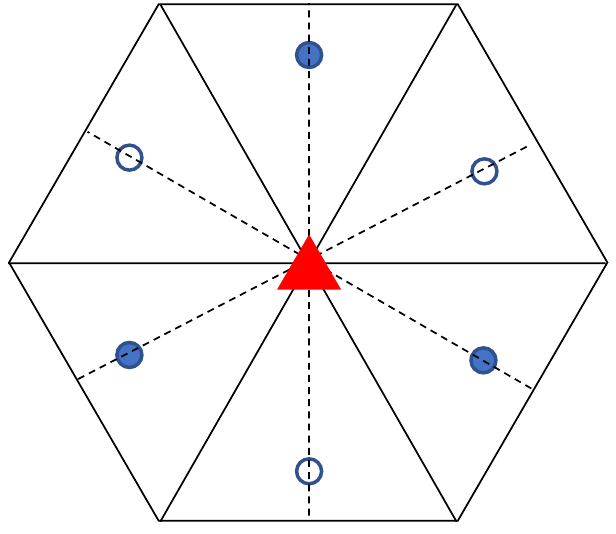}
		\caption{(Color online) Symmetry operations on a CrI$_3$ unit cell. Cr atoms sit at the corners of the hexagon, while the blue circles represent the I atoms located out-of-plane. Filled atoms are below the plane and empty circles are above the plane. The red triangle at the center of the hexagon represents C$_3$ rotation. Solid lines joining the vertices represent $C_2$ rotation axis, and dotted lines correspond to reflexion planes $\sigma_d$. Additionally, we find that $S_6 = \sigma_h C_6$ and inversion $i$ are also symmetries of the lattice.}
		\label{fig:D3d_ops}
	\end{center}
\end{figure}

Now we perform a group theory analysis on monolayer and bilayer CrI$_3$. Our goal is to determine the properties of the $\Gamma$ point phonons such as irreducible representations, lattice displacements, Raman and infrared activity and non-linear phonon coupling. For this, we employ GTPack \cite{gtpack1,gtpack2}, {\small ISOTROPY}~\cite{hatch2003s}, and the Bilbao Crystallographic Server~\cite{kroumova2003}. 

\subsubsection{Infrared and Raman active modes}
Ab-initio studies of the Raman spectrum on monolayer CrI$_3$ have postulated that the space group is $R\bar 3 2 m$ (No. 166)~\cite{larson2018}. However, more recent Raman experiments have identified the structure to belong to the $p \bar 3 1 m	$ ($D^1_{3d}$) double space group. 
\begin{figure*}[t]
	\begin{center}
		\includegraphics[width=17.0cm]{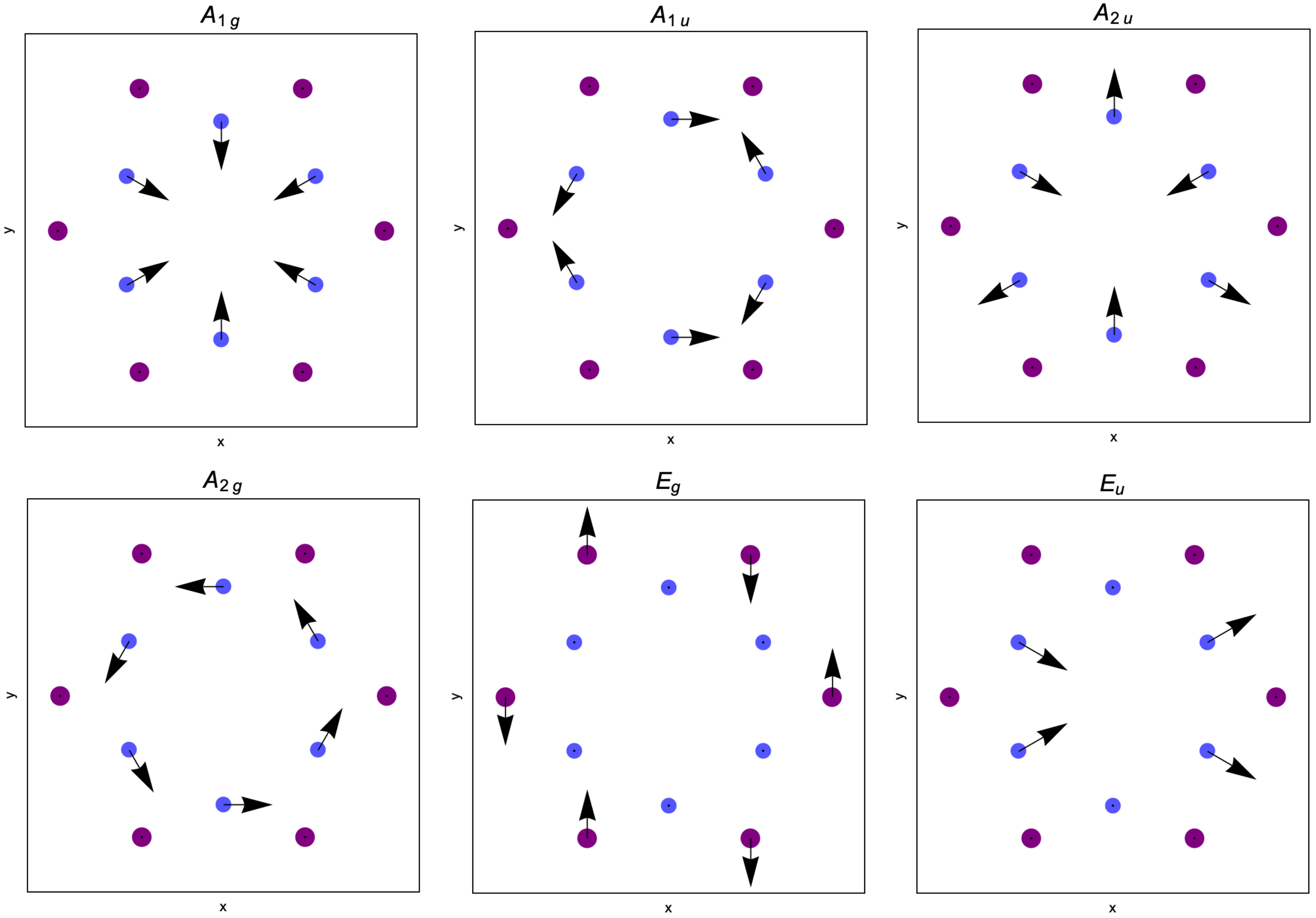}
		\caption{(Color online) CrI$_3$ vibrational modes projected onto the $xy$-plane for clarity. The one-dimensional representations ($A_j$) have one partner. The modes of the two-dimensional representations $E_j$ have two partners each which can be constructed by orthogonality with the shown modes.}
		\label{fig:vib_modes}
	\end{center}
\end{figure*}

In Fig. (\ref{fig:D3d_ops}) we show a diagram with the point group operations we identified in the lattice shown in monolayer CrI$_3$. The point group is found to be $D_{3d}$. The character table for the point group $D_{3d}$ is shown in panel (b)~\cite{GroupTheoryDress2008}. We start determining the infrared and Raman active modes in this system. For this, we first calculate the equivalence representation $\Gamma^{equiv}$ keeping in mind that there are 8 atoms per unit cell, two Cr atoms and six I atoms. $\Gamma^{equiv} $ is given in Table \ref{tab:equiv}.

\begin{table}[h]
\begin{tabular}{lllllll}
\cline{1-7}
  & $E$ & $C_3$ & $C_2$ & $i$ & $S_6$ & $\sigma_d$ \\ \cline{1-7}
$\Gamma^{equiv}$ & 8 & 2  & 2  & 0 & 0  & 2 
\end{tabular}
\caption{Equivalence representation.}
\label{tab:equiv}
\end{table}

\begin{table}[b]
\begin{tabular}{|c|c|c|c|c|c|c|c|c|}
\hline
 D$_{3d}$ & E & 2C$_3$ & 3C$_2$ & i & 2S$_6$ & 3 $\sigma_d $&  &  \\ \hline
 A$_{1g}$ & 1 &  1&  1&  1&  1& 1 &  &  $x^2+y^2, z^2$ \\ \hline
 A$_{2g} $& 1 &  1& -1&  1&  1& -1 & I$_z$ &  \\ \hline
 E$_g$&  2& -1 & 0 & 2 & -1 & 0 & (I$_x$,I$_y$)  & $(x^2-y^2,xy)$, $(yz,zx)$ \\ \hline
A$_{1u}$ &  1&  1&  1& -1 & -1 & -1 &  &  \\ \hline
A$_{2u}$ & 1 &  1& -1 & -1 & -1 &  1 &  z &  \\ \hline
E$_u$ & 2 & -1 & 0 &  -2& 1 & 0 & $(x,y)$ &  \\ \hline
\end{tabular}
\caption{D$_{3d}$ point group character table.}
\label{tab:d3d}
\end{table}
Using the decomposition theorem~\cite{GroupTheoryDress2008} we find

\begin{equation}
 \Gamma^{equiv} = 2A_{1g} \oplus E_g \oplus A_{1u} \oplus A_{2u} \oplus E_u. 
\end{equation}
 
 In this point group, the representation of the vector is $\Gamma_{vec} = E_u \oplus A_{2u}$. Then, the representation of the lattice vibrations is
 
\begin{align}\nonumber
\Gamma_{latt. vib.} & = \Gamma^{equiv} \otimes \Gamma_{vec} \\ & = 2A_{1g} \oplus 2A_{2g} \oplus 4E_g \oplus A_{1u} \oplus 3 A_{2u} \oplus 4 E_u.
\end{align}

From the character table, we can conclude that monolayer CrI$_3$ has six Raman active modes with representations $A_{1g}$, and $E_g$. Two frequencies are non-degenarate and four are doubly-degenerate. The two $A_{1g}$ modes correspond to ``breathing'' modes, where the lattice expands and contracts preserving all the symmetries. One of them is in-plane and the other one is out-of-plane.  Modes that transform as $A_{2u}$, and $E_u$ are infrared active. This results can be corroborated with the Bilbao Crystallography Server, and are consistent with the exiting literature~\cite{mijin2018}. The vibration eigenvectors for a given representation $\Gamma$ can be obtained using projection operators $P^{(\Gamma)} $ in the displacement representation as ~\cite{GroupTheoryDress2008}

\begin{equation}
Q^{\Gamma} = P^{(\Gamma)} \otimes \zeta,
\end{equation}
where $\zeta={x_1, y_1, z_1, \dots, x_N, y_N, z_N}$ is an arbitrary vector of dimension $3N$, and $N$ is the number of atoms in the unit cell. Fig. \ref{fig:vib_modes} shows some of the monolayer CrI$_3$ vibrational modes, projected onto the $xy$-plane for clarity. 

\section{Bilayer chromium triiodide group theory aspects.}
\label{app:group_bilayer}

In this section, we show explicitly the character table for the relevant point group, and the transformation from the conventional to the primitive unit cell. 

The transformation is given by

\begin{equation}
\mathcal T =
\begin{pmatrix}
1/2 & 1/2 & 0 \\
-1/2 & 1/2 & 0 \\
0 & 0 & 1
\end{pmatrix} + 
\begin{pmatrix}
1/2  \\
1/2  \\
0 
\end{pmatrix}.
\end{equation}

The character table for the point group C$_{2h}$ is

\begin{table}[h]
\begin{tabular}{|l|l|l|l|l|l|l|}
\hline
$C_{2h}$ & $E$ & $C_2$ & i  & $\sigma_h$ &              &                            \\ \hline
$A_g$    & 1   & 1     & 1  & 1          & $I_z$        & $x^2$, $y^2$, $ z^2$, $xy$ \\ \hline
$B_g$    & 1   & -1    & 1  & -1         & $I_x$, $I_y$ & $xz$, $yz$                 \\ \hline
$A_u$    & 1   & 1     & -1 & -1         & $z$          &                            \\ \hline
$B_u$    & 1   & -1    & -1 & 1          & $x,y$        &                            \\ \hline
\end{tabular}
\caption{C$_{2h}$ point group  character table~\cite{GroupTheoryDress2008}.}
\label{tab:c2h}
\end{table}

\end{document}